\documentclass[reprint,amsmath,amssymb,aip,jcp,floatfix,
               superscriptaddress]{revtex4-1}
\usepackage{times,color,graphicx,multirow,placeins}
\usepackage[utf8]{inputenc}
\usepackage[colorlinks=true,urlcolor=blue,linkcolor=blue,
            citecolor=blue]{hyperref}
\usepackage{amsmath} 
\newcommand{\bibtitle}[1]{\textit{#1},}

\begin{document}

\title{X marks the spot:\@ accurate energies from intersecting
extrapolations of continuum quantum Monte Carlo data}

\author{Seyed Mohammadreza Hosseini}
  \affiliation{Max-Planck Institute for Solid State Research,
               Heisenbergstr.\ 1, 70569 Stuttgart, Germany}
\author{Ali Alavi}
  \affiliation{Max-Planck Institute for Solid State Research,
               Heisenbergstr.\ 1, 70569 Stuttgart, Germany}
  \affiliation{Yusuf Hamied Department of Chemistry,
               University of Cambridge,
               Lensfield Road, Cambridge CB2 1EW, UK}
\author{Pablo L\'opez R\'ios}
  \email{p.lopez.rios@fkf.mpg.de}
  \affiliation{Max-Planck Institute for Solid State Research,
               Heisenbergstr.\ 1, 70569 Stuttgart, Germany}

\begin{abstract}
We explore the application of an extrapolative method that yields very
accurate total and relative energies from variational and diffusion
quantum Monte Carlo (VMC and DMC) results.
For a trial wave function consisting of a small configuration
interaction (CI) wave function obtained from full CI quantum Monte
Carlo and reoptimized in the presence of a Jastrow factor and an
optional backflow transformation, we find that the VMC and DMC
energies are smooth functions of the sum of the squared coefficients
of the initial CI wave function, and that quadratic extrapolations of
the non-backflow VMC and backflow DMC energies intersect within
uncertainty of the exact total energy.
With adequate statistical treatment of quasi-random fluctuations, the
\textit{e{\underline x}trapolate and inter{\underline s}ect with
{\underline p}olynomials of {\underline o}rder {\underline t}wo}
(\textsc{xspot}) method is shown to yield results in agreement with
benchmark-quality total and relative energies for the C$_2$, N$_2$,
CO$_2$, and H$_2$O molecules, as well as for the C$_2$ molecule in its
first electronic singlet excited state, using only small CI expansion
sizes.
\end{abstract}

\maketitle

\section{Introduction}

Quantum Monte Carlo (QMC) methods are a broad family of stochastic
wave-function-based techniques that accurately approximate the
solution of the Schr\"odinger equation of an electronic system.
The variational quantum Monte Carlo (VMC) method
\cite{McMillan_he_1965, Needs_casino_2020} obtains expectation values
corresponding to an analytic trial wave function $\Psi_{\rm T}({\bf
R})$ in real space and provides a framework for optimizing wave
function parameters, \cite{Toulouse_emin_2007, Umrigar_emin_2007} such
as those in the multideterminant-Jastrow-backflow form,
\begin{equation}
  \label{eq:mdet_jastrow_backflow}
  \Psi_{\rm T}({\bf R}) = e^{J({\bf R})}
     \sum_I^{M} c_I D_I[{\bf X}({\bf R})] \;,
\end{equation}
where $\{D_I\}$ are $M$ Slater determinants, $e^{J({\bf R})}$ is a
Jastrow correlation factor, \cite{Drummond_jastrow_2004,
LopezRios_jastrow_2012} and ${\bf X}({\bf R})$ are
backflow-transformed electronic coordinates.
\cite{LopezRios_backflow_2006}
Diffusion quantum Monte Carlo (DMC) \cite{Ceperley_DMC_1980,
Needs_casino_2020} is a real-space projection method which recovers
the lowest-energy solution $\Phi({\bf R})$ of the Schr\"odinger
equation compatible with the fixed-node condition that $\Phi({\bf R)}
\Psi_{\rm T}({\bf R})$ be nonnegative everywhere.
We refer to the VMC and DMC methods collectively as continuum QMC
(cQMC) methods.

In the configuration interaction (CI) ansatz the solution of the
Schr\"odinger equation is expressed as
\begin{equation}
  \label{eq:CI_ansatz}
  |\Psi\rangle = \sum_I c_I |D_I\rangle \;,
\end{equation}
where $\{|D_I\rangle\}$ are all possible Slater determinants that can
be constructed in a given orbital basis.
Equation \ref{eq:CI_ansatz} is exact in the limit of an infinite
expansion using a complete basis set, but in practical methods finite
expansions and bases are used.
Full CI quantum Monte Carlo (FCIQMC) \cite{Booth_fciqmc_2009,
Cleland_initiator_2010} is a second-quantized projection technique in
which random walkers sample the discrete Hilbert space defined by
$\{|D_I\rangle\}$ in order to determine approximate values of the CI
coefficients $\{c_I\}$.

The nominal computational cost of cQMC calculations scales
polynomially with system size $N$, typically as $N^2$--$N^4$, and the
quality of the resulting energies depends on the accuracy of the trial
wave function for VMC, and on the accurate location of its nodes for
DMC.
The cQMC methods excel at describing explicit dynamic and long-ranged
correlations, but the error incurred by the fixed-node approximation
is often significant.
By contrast, FCIQMC is formally an exponentially-scaling method which
trivially captures static correlations, but requires a very large
number of walkers to provide a good description of dynamic
correlations.
The complementary nature of the strengths of cQMC and FCIQMC makes
combining these methods highly desirable.
Several ways of combining cQMC and FCIQMC have been presented in the
literature, such as using DMC to assist in the extrapolation to the
thermodynamic limit of FCIQMC energies of the electron gas,
\cite{Ruggeri_ueg_2018} or using VMC-optimized Jastrow factors in
FCIQMC with the transcorrelated method.
\cite{Cohen_fciqmc_jastrow_2019, Haupt_optimize_jastrow_2023}
Here we shall focus on the use of selected CI wave functions
generated with FCIQMC to construct multideterminantal trial wave
functions for cQMC calculations.

Multideterminant expansions have been used for decades in cQMC
calculations of atomic and molecular systems, including ground-state
energy calculations, \cite{Filippi_diatomics_1996, Brown_atoms_2007,
Seth_atoms_2011, Petruzielo_mdet_2012, Morales_mdet_2012,
Giner_mdet_2016} excitation energies, \cite{Scemama_ex_mdet_2018,
Scemama_ex_mdet_2019, Dash_mdet_2019, Dash_mdet_2021} and geometry
optimizations.  \cite{Dash_mdet_2018, Dash_mdet_2019}
The use of truncated CI expansions in cQMC presents the problem that
no reliable criteria exist to truncate wave functions for different
systems in a consistent manner, resulting in energy differences of
questionable accuracy.
One possible approach is to use extremely large multideterminantal
wave functions, \cite{Giner_mdet_2016, Scemama_mdet_2016,
Per_mdet_2017, Dash_mdet_2018, Scemama_ex_mdet_2018,
Scemama_ex_mdet_2019, Dash_mdet_2019, Scemama_nod_mdet_2018,
Dash_mdet_2021} under the expectation that the fixed-node error in the
total energies will become smaller than the target error.
While algorithmic developments have vastly reduced the computational
cost associated with the use of multideterminantal wave functions in
cQMC, \cite{Nukala_mdet_2009, Clark_mdet_2011, Weerasinghe_mdet_2014,
Filippi_mdet_2016, Scemama_mdet_2016} this remains an expensive
choice.
Using trial wave functions without a Jastrow factor reduces the
nominal computational burden \cite{Scemama_ex_mdet_2018,
Scemama_ex_mdet_2019, Scemama_nod_mdet_2018} at the cost of losing the
accurate, compact description of dynamic correlation afforded by
fully-optimized trial wave functions.
By including explicit correlations, in the present paper we are able
to explore the use of relatively small multideterminantal wave
functions to perform an extrapolation of the cQMC total energy to the
full-CI, complete orbital-basis limit.
We test our method on a variety of molecular systems, obtaining total
and relative energies within uncertainty of benchmark-quality results
from the literature.

The rest of this paper is structured as follows.
In Section \ref{sec:method} we present the methodological details of
our extrapolation method, which we illustrate with calculations of
the carbon dimer and the water molecule.
We then apply our method to several atomic and molecular systems, and
we report the results in Section \ref{sec:results}.
Our conclusions and outlook are presented in Section
\ref{sec:conclusions}.
Hartree atomic units ($\hbar=|e|=m_e=4\pi\epsilon_0=1$) are used
throughout; the uncertainties and error bars we report refer to
standard 68.3\% (one-sigma) confidence intervals except when
explicitly noted otherwise.

\section{Methodology}
\label{sec:method}

Let $M_{\rm gen}$ be the number of determinants occupied at a given
point in an equilibrated FCIQMC calculation, representing the CI wave
function
\begin{equation}
  \label{eq:truncated_CI}
  |\Psi_{M_{\rm gen}}\rangle = \sum_{I=1}^{M_{\rm gen}}
                               c_I |D_I\rangle \;,
\end{equation}
where $c_I$ is obtained as the sum of the signed weights of the
walkers occupying the $I$th determinant.
The values of the first few coefficients $\{c_I\}_{I\ll M_{\rm gen}}$
converge relatively quickly in FCIQMC calculations and can be expected
to be reasonably close to their full CI (FCI) values.
This makes FCIQMC an ideal method for quickly generating good-quality
selected CI wave functions of moderate sizes -- studying the
suitability of other CI solvers for this purpose is beyond the scope
of this paper.

Let us consider the wave function obtained by truncating Eq.\
\ref{eq:truncated_CI} to size $M\ll M_{\rm gen}$.
The sum of the squares of the coefficients of the resulting wave
function relative to that at size $M_{\rm gen}$ is
\begin{equation}
  \label{eq:w_def}
  w^2 = \frac{\sum_{I=1}^M c_I^2}
             {\sum_{I=1}^{M_{\rm gen}} c_I^2} \;,
\end{equation}
which goes to $1$ as $M\to M_{\rm gen}$.
This CI wave function of size $M$ can be combined with a Jastrow
factor, and optionally with a backflow transformation, to produce a
multideterminant-Jastrow(-backflow) trial wave function for cQMC, as
given in Eq.\ \ref{eq:mdet_jastrow_backflow}.
The wave function parameters can be (re-)optimized in the context of
VMC, producing a trial wave function with which to compute VMC and DMC
energies.
Repeating this procedure by truncating the original CI wave function
to different sizes yields a set of VMC and DMC energies that can be
plotted as a function of $w$.

\begin{figure}[!hbt]
  \begin{center}
    \includegraphics[width=\columnwidth]{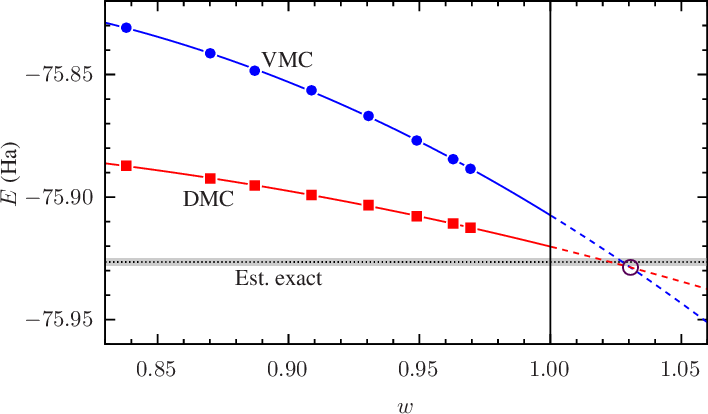}
    \caption{
      VMC and DMC total ground-state energy of the carbon dimer using
      multideterminant-Jastrow trial wave functions as a function of
      $w$, using Hartree-Fock orbitals expanded in the cc-pCVTZ basis
      set.
      Quadratic fits to the data are extended beyond $w=1$ to show
      their intersection, which is in good agreement with the
      estimated exact nonrelativistic total energy of the system.
      \cite{Bytautas_diatomic_2005}
    }
    \label{fig:c2_demo}
  \end{center}
\end{figure}

Plots of this kind, albeit using other CI solvers, can be found in the
literature; see Fig.\ 3 of Ref.\ \onlinecite{Umrigar_emin_2007} or
Fig.\ 4 of Ref.\ \onlinecite{Clark_mdet_2011}, for example.
The present work is in fact inspired by the observation that the VMC
and DMC curves in these plots appear to be smooth and would seem to be
about to intersect just off the right-hand side of the graph.
In Fig.\ \ref{fig:c2_demo} we plot the VMC and DMC energies we obtain
for the ground state of the all-electron C$_2$ molecule, see Table
\ref{table:systems}, using Hartree-Fock orbitals expanded in the
cc-pCVTZ basis set, \cite{Dunning_cc-pvxz_1989, Woon_cc-pcvxz_1995}
along with quadratic fits to the data of the form
\begin{equation}
  \label{eq:Ew_model}
  E(w) = a + bw + cw^2 \;,
\end{equation}
where $a$, $b$, and $c$ are fit parameters.
In this simple example, the fits to the VMC and DMC data intersect at
$w=1.031$, corresponding to a total energy of $-75.9287$ Ha, not far
off the exact nonrelativistic total energy estimate of $-75.9265$ Ha
given in Ref.\ \onlinecite{Bytautas_diatomic_2005}.
We refer to this way of estimating the total energy of a system as the
\textit{e{\underline x}trapolate and inter{\underline s}ect with
{\underline p}olynomials of {\underline o}rder {\underline t}wo}
(\textsc{xspot}) method.

In what follows we develop the methodology to enable the application
of the \textsc{xspot} method in practice using as test systems the C,
N, and O atoms, the ground-state C$_2$, N$_2$, H$_2$O, and CO$_2$
molecules, and the C$_2$ molecule in its lowest-lying singlet
electronic excited state, which we refer to simply as
${{\textrm{C}}_2}^*$.
These atoms and molecules are simulated as all-electron, both in the
sense that no effective-core potentials are used and that excitations
from ``core'' orbitals are allowed in the CI wave function.
In Table \ref{table:systems} we give the states and geometries we have
used for these systems.
\begin{table}[ht!]
  \begin{tabular}{cll}
    \multicolumn{1}{c}{System    } &
    \multicolumn{1}{c}{Geometry \cite{Bytautas_diatomic_2005,
                                      Feller_molecules_2008} } &
    \multicolumn{1}{c}{State     } \\
    \hline
    \hline
    C      & & $^3P$                                         \\[0cm]
    N      & & $^4S$                                         \\[0cm]
    O      & & $^3P$                                         \\[0cm]
    C$_2$  & $r_{\rm CC} = 1.2425~{\rm \AA}$ & $^1\Sigma_g^+$\\[0cm]
    ${{\textrm{C}}_2}^*$&
             $r_{\rm CC} = 1.2425~{\rm \AA}$ & $B^1\Delta_g$ \\[0.05cm]
    \parbox[t]{8mm}{\multirow{2}{*}{H$_2$O}}
           & $r_{\rm OH} = 0.9572~{\rm \AA}$ &
    \parbox[t]{8mm}{\multirow{2}{*}{$^1A_1$}}                 \\
           & $\angle_{\rm HOH}=104.5^\circ$  &                \\[0.05cm]
    N$_2$  & $r_{\rm NN} = 1.0977~{\rm \AA}$ & $^1\Sigma_g^+$ \\[0.05cm]
    \parbox[t]{8mm}{\multirow{2}{*}{CO$_2$}}
           & $r_{\rm CO} = 1.1600~{\rm \AA}$ &
    \parbox[t]{8mm}{\multirow{2}{*}{$^1\Sigma_g^+$}}          \\
           & $\angle_{\rm OCO}=180^\circ$    &                \\
    \hline
  \end{tabular}
  \caption{
    Atoms and molecules considered in this work, along with their
    electronic states and geometries.
  }
  \label{table:systems}
\end{table}

\subsection{Theoretical justification}
\label{sec:justification}

The extrapolation shown in Fig.\ \ref{fig:c2_demo} might seem
simplistic from a quantum chemical perspective, given that all
calculations involved have been performed with the same, finite
orbital basis, so one would expect an orbital-basis dependent result
which should itself be extrapolated to the complete-basis limit.
For instance, the FCIQMC energy tends to a basis-set dependent FCI
limit as the number of walkers tends to infinity, and this must in
turn be extrapolated to the basis-set limit in order to obtain the
exact energy of the system.

In what follows we will conceptually combine the choice of molecular
orbitals (e.g., Hartree-Fock, natural orbitals, \ldots) with the
choice of basis set (e.g., cc-pCVDZ, cc-aug-pVTZ, \ldots), so we shall
discuss the completeness of the (molecular) orbital basis instead of
that of the basis set alone to emphasize this point.

The \textsc{xspot} extrapolation procedure can be easily justified in
the hypothetical case of using an infinite, ``complete'' orbital
basis.
The FCIQMC energy with this ``complete'' orbital basis would tend to
the exact total energy of the system $E_0$ in the infinite
walker-number limit, and the sum of the squared CI coefficients would
also tend to that of the exact wave function, $w_0$.
The exact wave function has no dynamic correlation left to recover, so
the Jastrow factor and backflow displacement in the cQMC trial wave
function would optimize to zero, and the VMC and DMC energies would
both coincide with $E_0$.
At finite expansion sizes, $w < w_0$, the VMC and DMC methods yield
variational energies satisfying $E_{\rm VMC} \geq E_{\rm DMC} \geq
E_0$, which, assuming these to be smooth functions of $w$, validates
the \textsc{xspot} method with the ``complete'' orbital basis.

We note that in a truncated CI wave function, the infinite,
``complete'' orbital basis is effectively finite, since a finite
number of determinants can only contain a finite number
of distinct orbitals.
Conversely, a sufficiently small selected CI wave function with a
finite orbital basis is indistinguishable from a CI wave function of
the same size with the ``complete'' orbital basis -- assuming the
finite basis contains the first few orbitals in the ``complete''
orbital basis.

As the orbitals in a finite basis get used up, the cQMC energies can
be expected to plateau as a function of $w$ as they tend to their
orbital-basis dependent limit.
We refer to this phenomenon as ``orbital-basis exhaustion'', and to
the onset of this plateau as the exhaustion limit $w_{\rm exh.}$.
Note that orbital bases such as natural orbitals can be constructed so
as to compactly describe the system with fewer orbitals, which has the
side effect of reducing the value of $w_{\rm exh.}$.
We discuss this aspect further in Section \ref{sec:basis}.

As a proxy for the degree of orbital-basis exhaustion, in Table
\ref{table:c2_exhaustion} we show the fraction of orbitals used in CI
wave functions of the same size for Hartree-Fock orbitals expanded in
four different basis sets in the cc-pV$x$Z and cc-pCV$x$Z families
\cite{Dunning_cc-pvxz_1989, Woon_cc-pcvxz_1995} for the all-electron
carbon dimer.
\begin{table}[htbp]
  \centering
  \begin{tabular}{lr@{$~=~$}l}
    \multicolumn{1}{c}{Basis set}     &
    \multicolumn{2}{c}{Orbitals used} \\
    \hline \hline
    cc-pVDZ  & $28$/$28$ & $100\%$ \\
    cc-pCVDZ & $32$/$36$ & $89\% $ \\
    cc-pVTZ  & $38$/$60$ & $82\% $ \\
    cc-pCVTZ & $40$/$86$ & $47\% $ \\
    \hline
  \end{tabular}
  \caption{
    Fraction of spatial orbitals in the Hartree-Fock orbital basis
    expanded in each of four basis sets that appear in the first 300
    configuration state functions of the full CI wave function for the
    ground state of the carbon dimer.
  }
  \label{table:c2_exhaustion}
\end{table}
Based on these numbers, we use the cc-pCVTZ basis throughout this
paper to ensure we have enough leeway to increase the multideterminant
wave function size before hitting the exhaustion limit.
We provide an \textit{a posteriori} assessment of this choice in
Section \ref{sec:results}.

Finite-orbital-basis FCIQMC and cQMC calculations performed at
$w<w_{\rm exh.}$ behave as if one were using the ``complete'' orbital
basis.
Therefore it is legitimate to expect that the extrapolation of
quadratic fits to these VMC and DMC data intersect at $w=w_0$ and
$E=E_0$, provided that the VMC and DMC energies are smooth functions
of $w$ representable by a second-order polynomial for $w_{\rm h.o.} <
w < w_{\rm exh.}$, where $w_{\rm h.o.}$ is a threshold below which
higher-order polynomials would be needed.

Note that the initial FCIQMC wave function with $M_{\rm gen}$
determinants is not required to be below the exhaustion limit since it
simply serves to construct selected CI wave functions of size $M\ll
M_{\rm gen}$, which \emph{are} required to be below the exhaustion
limit, and to define the arbitrary point at which $w=1$ in the plots;
$w=1$ has no special significance in this method.

In our calculations we choose CI wave function sizes so that the
points are more or less evenly spaced in the $w$ axis, and we make
sure that different points correspond to wave functions containing a
different number of distinct spatial orbitals so as to capture the
effect of simultaneously growing the CI expansion and the orbital
basis.

\subsection{Obtaining statistically meaningful results}
\label{sec:method_stats}

\begin{figure}[!hbt]
  \begin{center}
    \includegraphics[width=\columnwidth]{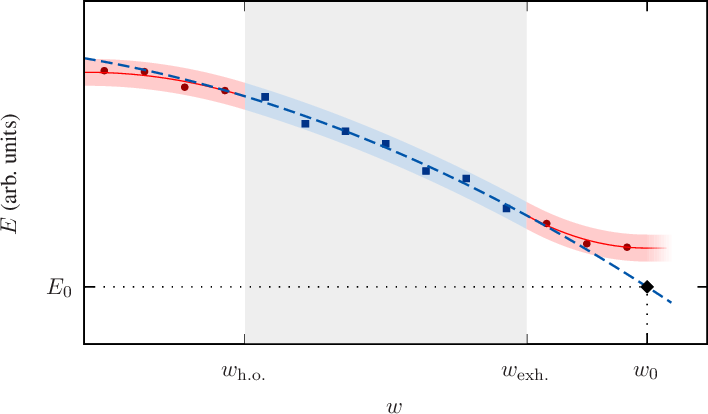}
    \caption{
      Illustration of the expected behavior of cQMC energies as a
      function of $w$.
      The cQMC energies (circles and squares) deviate from
      the underlying smooth trend (lines) by a quasirandom amount
      (of amplitude represented by the width of the shaded area around
      the lines).
      The smooth trend can be represented by a quadratic function,
      $E(w)$ (dashed line), for $w_{\rm h.o.}<w<w_{\rm exh.}$ (shaded
      middle region), while for $w<w_{\rm h.o.}$ higher-order
      contributions become important, and for $w>w_{\rm exh.}$
      orbital-basis exhaustion sets in.
      At the value of $w$ corresponding to the exact wave function
      the quadratic function gives the exact energy, $E_0=E(w_0)$.
    }
    \label{fig:model_cartoon}
  \end{center}
\end{figure}

In reality cQMC energies do not exactly follow smooth curves, but it
is reasonable to assume that a smooth underlying trend $E(w)$ exists,
and that the cQMC energy $E_i$ deviates from it by a a quasi-random
amount $q_i$.
Considering also the statistical uncertainty $\Delta E_i$, the $i$th
point in a set of cQMC energies can then be modelled as
\begin{equation}
  \label{eq:qrandom_model}
  E_i = E(w_i) + q_i + \zeta_i \Delta E_i \;,
\end{equation}
where
$\zeta_i$ is a random number drawn from the standard normal
distribution.
In order to make this generic model for the quasi-random fluctuations
useful in practice, we make the approximation that $q_i=\xi_i\alpha$,
where $\xi_i$ is a random number drawn from the standard normal
distribution and $\alpha$ is a constant amplitude independent of $w$.
In Fig.\ \ref{fig:model_cartoon} we illustrate our model of cQMC
energy data as a function of $w$.

We estimate the value of $\alpha$ by performing a preliminary
least-squares fit to the bare data, ${\tilde E}(w)$, and evaluating
\begin{equation}
  \label{eq:alpha2}
  \alpha^2 =
    \max\left(0,
      \left\langle
          \left[ E_i - {\tilde E}(w_i) \right]^2
      \right\rangle
      -
      \left\langle
        \left(\Delta {E_i}\right)^2
      \right\rangle
    \right) \;,
\end{equation}
i.e., we obtain $\alpha$ as the root-mean-square deviation
of the data from the fit value \textit{not accounted for by
statistical uncertainty alone}.
For this procedure to produce a meaningful result, the number of data
points in each curve must be significantly greater than the number of
parameters in the quadratic fit function; we use at least $7$ data
points for all fits reported in this paper.

\begin{figure}[!hbt]
  \centering
  \includegraphics[width=\columnwidth]{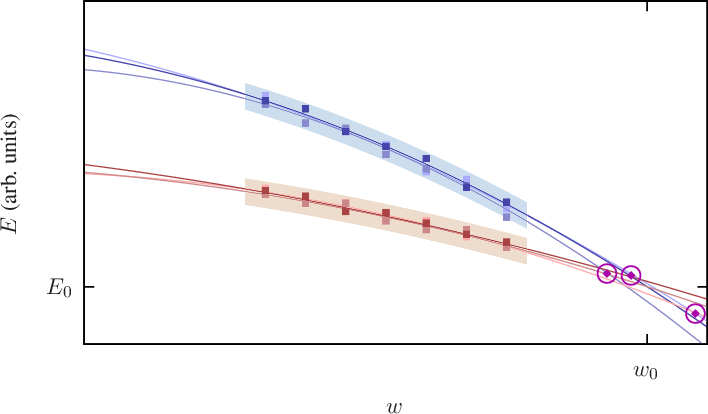}
  \caption{
    Illustration of the Monte Carlo resampling scheme used to
    compute statistics on the intersection between two curves.
    For each curve, having obtained an estimate of $\alpha$ (width
    of the shaded region) from the original energy data (not shown),
    we create a synthetic instance of the dataset by shifting the
    original points by a random amount proportional to
    $\sqrt{\alpha^2 + \Delta {E_i}^2}$ (squares of same color
    saturation), perform a quadratic fit (line), and find the
    intersection between both fits (circled diamond).
    This process is repeated over the random instances (three shown
    in the illustration), from which statistics on the intersection
    are obtained.
  }
  \label{fig:resample_cartoon}
\end{figure}

In order to account for the statistical uncertainty and quasi-random
fluctuations in the \textsc{xspot} method, we use a Monte Carlo
resampling technique in which we generate $100\,000$ instances of each
VMC and DMC dataset in which a random amount proportional to
$\sqrt{\alpha^2 + \Delta {E_i}^2}$ is added to the original energy
values.
We then perform fits to these shifted data and find the intersection
point for each such instance, and obtain the final result by averaging
over instances; see Fig.\ \ref{fig:resample_cartoon} for an
illustration of this process.
This procedure provides meaningful uncertainties on the intersection
energies which account for both the cQMC statistical uncertainty and
quasi-random deviations from the smooth trend.

\begin{figure}[!hbt]
  \centering
  \includegraphics[width=\columnwidth]{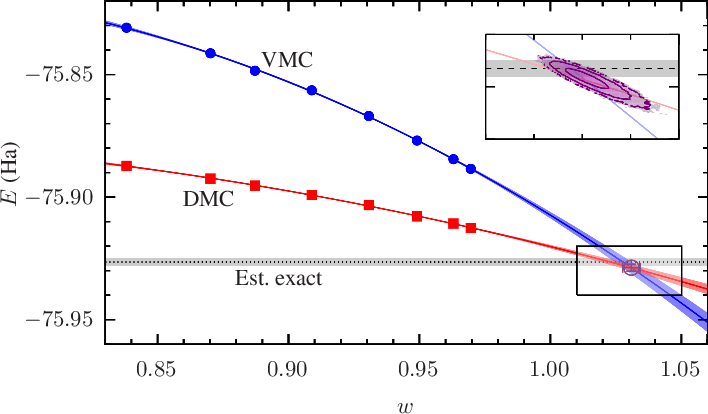}
  \caption{
    VMC and DMC energies of the ground-state C$_2$ molecule as a
    function of $w$, as shown in Fig.\ \ref{fig:c2_demo}, using the
    full statistical treatment of the \textsc{xspot} method.
    Mean values of the fits to the data are shown as lines, and the
    translucent areas around them represent 95.5\% (two-sigma)
    confidence intervals.
    Also shown is the estimated exact nonrelativistic energy
    \cite{Bytautas_diatomic_2005} as a dotted line with a shaded area
    of $\pm 1$ kcal/mol around it, and the intersection point between
    the curves.
    The inset shows the statistical distribution of intersection
    points as a color map with overlaid contour curves.
  }
  \label{fig:c2_demo_2}
\end{figure}

We demonstrate the full statistical procedure of the \textsc{xspot}
method on multideterminant-Jastrow data for the carbon dimer in Fig.\
\ref{fig:c2_demo_2}.
Notice that the distribution of intersection points shown in the inset
of Fig.\ \ref{fig:c2_demo_2} has a tail extending towards low $E$ and
large $w$.
These tails become more problematic the more parallel the two
intersecting curves are, eventually preventing the evaluation of an
intersection point at all.
It is therefore important to try to apply the \textsc{xspot} method to
curves which are as close to perpendicular as possible.

\begin{figure}[!hbt]
  \centering
  \includegraphics[width=\columnwidth]{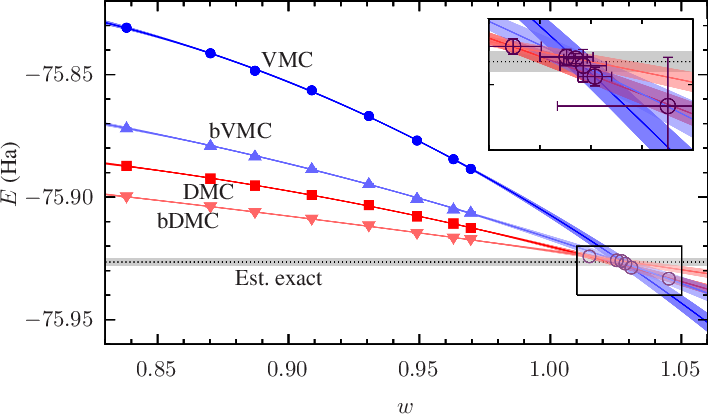}
  \caption{
    VMC, DMC, bVMC, and bDMC energies of the ground-state C$_2$
    molecule as a function of $w$.
    Mean values of the fits to the data are shown as lines, and the
    translucent areas around them represent 95.5\% (two-sigma)
    confidence intervals.
    Also shown is the estimated exact nonrelativistic energy
    \cite{Bytautas_diatomic_2005} as a dotted line with a shaded area
    of $\pm 1$ kcal/mol around it, and the intersection points between
    each of the six possible pairs of curves; error bars on these are
    only shown in the inset for clarity.
  }
  \label{fig:c2_demo_3}
\end{figure}

In Fig.\ \ref{fig:c2_demo_3} we include additional VMC and DMC data
using an inhomogeneous backflow transformation (``bVMC'' and ``bDMC'')
for the carbon dimer.
We list the intersections between pairs of curves in Table
\ref{table:c2_intersections}, all of which are within uncertainty of
each other.
\begin{table}[htbp]
  \centering
  \begin{tabular}{clcc}
    \multicolumn{1}{c}{Curves      } &
    \multicolumn{1}{c}{$w_0$       } &
    \multicolumn{1}{c}{$E_0$ (a.u.)} &
    \multicolumn{1}{c}{Miss (\%)   } \\
    \hline \hline
    VMC-DMC   & $1.031(3)$ & $-75.9288 \pm 0.0014$ & $0.00$ \\
    VMC-bVMC  & $1.028(5)$ & $-75.9271 \pm 0.0025$ & $0.00$ \\
    VMC-bDMC  & $1.027(2)$ & $-75.9260 \pm 0.0008$ & $0.00$ \\
    DMC-bVMC  & $1.05(2) $ & $-75.9333 \pm 0.0074$ & $2.75$ \\
    DMC-bDMC  & $1.015(6)$ & $-75.9242 \pm 0.0012$ & $0.00$ \\
    bVMC-bDMC & $1.025(5)$ & $-75.9258 \pm 0.0012$ & $0.00$ \\
    \hline
  \end{tabular}
  \caption{
    Location of all six pairwise intersections of the VMC, DMC,
    bVMC, and bDMC curves shown in Fig.\ \ref{fig:c2_demo_3} for
    the C$_2$ molecule.
    ``Missed intersections'' refer to random instances of the curves
    that do not intersect at $w>1$ in the Monte Carlo resampling
    procedure.
  }
  \label{table:c2_intersections}
\end{table}
The VMC and bDMC curves provide the best-resolved results, which is to
be expected since these curves intersect at the widest angle among all
pairs of curves, as can be seen in Fig.\ \ref{fig:c2_demo_3}.
By contrast, the DMC and bVMC curves intersect at a narrow angle and
incur a small but non-zero fraction of ``missed'' intersections, i.e.,
random instances of the data whose fits fail to intersect at $w>1$,
which signals the presence of heavy tails in the intersection
distribution, resulting in a large uncertainty on the intersection
energy.

For the four curves in Fig.\ \ref{fig:c2_demo_3}, the estimated
amplitude of the quasi-random fluctuations ranges from $\alpha=0.17$
to $0.41$~mHa.
Throughout this paper we converge the cQMC calculations so that the
uncertainties satisfy $\langle (\Delta E_i)^2 \rangle < \alpha^2$,
i.e., so that quasirandom fluctuations represent the main contribution
to the uncertainty on the fits and intersections; see the
supplementary material for the list of values of $\alpha$ obtained.
The statistical uncertainties on the cQMC energies can thus be
neglected for all practical purposes.

\subsection{Choice of orbital basis}
\label{sec:basis}

\begin{figure*}[!hbt]
  \centering
  \includegraphics[width=0.92\columnwidth]{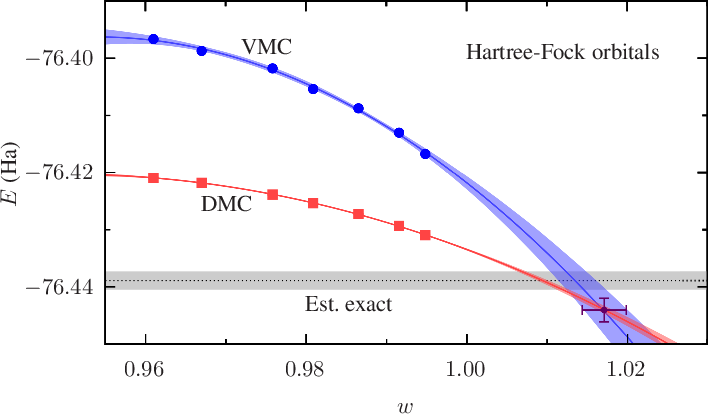} \quad
  \includegraphics[width=0.92\columnwidth]{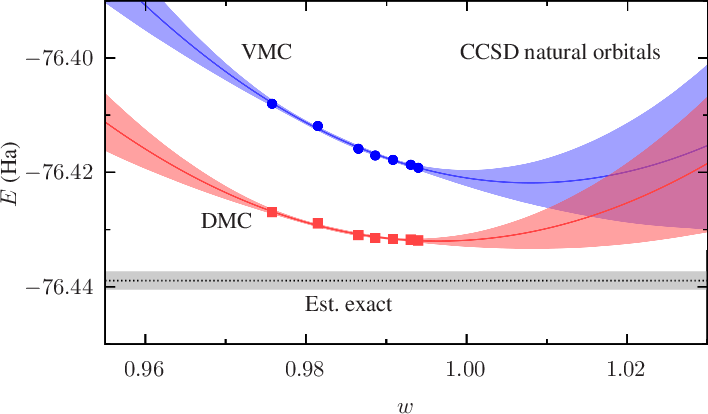}
  \caption{
    VMC and DMC energies of the H$_2$O molecule as a function of $w$,
    both using Hartree-Fock orbitals (left) and CCSD natural orbitals
    (right) expanded in the cc-pCVTZ basis set.
    In both cases the same numbers of CSFs are used.
    Mean values of the fits to the data are shown as lines, and the
    translucent areas around them represent 95.5\% (two-sigma)
    confidence intervals.
    Also shown is the estimated exact nonrelativistic energy
    \cite{Feller_molecules_2008} as a dotted line with a shaded area
    of $\pm 1$ kcal/mol around it, and the intersection point between
    the VMC and DMC curves in the left panel.
    The cQMC energies obtained with natural orbitals plateau with $w$,
    preventing the quadratic extrapolations from reaching the exact
    energy.
  }
  \label{fig:h2o_nat}
\end{figure*}

As alluded to in Section \ref{sec:justification}, the choice of
molecular orbitals plays a crucial role in the behavior of the cQMC
energies as a function of $w$, modifying the point $w_{\rm exh.}$ at
which the effects of exhaustion start to become noticeable.
In Fig.\ \ref{fig:h2o_nat} we demonstrate this for the H$_2$O molecule
by comparing the cQMC energies obtained using Hartree-Fock orbitals
expanded in the cc-pCVTZ basis set and natural orbitals expanded in
the same basis set constructed so as to diagonalize the one-body
density matrix in coupled cluster singles and doubles (CCSD).
While CCSD natural orbitals produce lower cQMC energies and correspond
to larger values of $w$ at fixed expansion sizes, the cQMC energies we
obtain using Hartree-Fock orbitals follow the quadratic trend
throughout the whole $w$ range considered, while those obtained with
natural orbitals plateau very early on, preventing their meaningful
extrapolation.

\section{Results and discussion}
\label{sec:results}

\subsection{Calculation details}

In our calculations we use Hartree-Fock orbitals expanded in the
cc-pCVTZ Gaussian basis set \cite{Dunning_cc-pvxz_1989,
Woon_cc-pcvxz_1995} obtained using \textsc{molpro}.  \cite{molpro}
We perform a small-scale FCIQMC calculation using the \textsc{neci}
package \cite{Guther_neci_2020} with configuration state functions
(CSFs) instead of determinants as walker sites,
\cite{Dobrautz_GUGA_2019} which reduces the number of FCIQMC walkers
required to accurately represent the wave function; note that the use
of CSFs is not a requirement of the \textsc{xspot} method.
The FCIQMC population is grown to $10^6$ walkers and equilibrated, and
the coefficients of the $M_{\rm gen}=10\,000$ most-occupied CSFs are
recorded from a population snapshot.
From this information we build CI expansions with the $M$ CSFs with
largest absolute coefficients, where $M \leq 1500 \ll M_{\rm gen}$.
In our cQMC calculations the orbitals are corrected to obey the
electron-nucleus cusp condition. \cite{Ma_cusp_2005}

The CSF coefficients are reoptimized in the presence of a Jastrow
factor of the Drummond-Towler-Needs form, \cite{Drummond_jastrow_2004,
LopezRios_jastrow_2012} and of an optional inhomogeneous backflow
transformation including electron-electron, electron-nucleus, and
electron-electron-nucleus terms. \cite{LopezRios_backflow_2006}
We do not optimize any of the parameters in the molecular orbitals,
which provide degrees of freedom that overlap significantly with those
in the backflow transformation.
Note that even though CSFs are used, the presence of the Jastrow
factor and of the backflow transformation prevents the cQMC trial wave
function from formally being an exact spin state.
\cite{Huang_spin-contamination_1998}
We optimize our wave function parameters using linear least-squares
energy minimization \cite{Toulouse_emin_2007, Umrigar_emin_2007} with
$10^6$ statistically independent VMC-generated electronic
configurations, a number large enough that the optimized cQMC energy
can be assumed to lie reasonably close to its variational minimum;
note that any remaining optimization error can be considered to be
absorbed into the quasirandom error.

The resulting trial wave function is then used to run two DMC
calculations with time steps 0.001 and 0.004 a.u.\@ and target
populations of 2048 and 512 configurations, respectively, except for
bDMC runs on CO$_2$ at $500$ and $1000$ CSFs for which we use 65536
and 16384 configurations.
These energies are then linearly extrapolated to the zero time-step,
infinite-population limit.  \cite{Needs_casino_2020,
Lee_strategies_2011}

We use the \textsc{casino} package \cite{Needs_casino_2020} to run the
cQMC calculations, and use multi-determinant compression
\cite{Weerasinghe_mdet_2014} to reduce the computational expense of
evaluating the trial wave function.
We perform the fits to the data and find their intersections using our
custom \textsc{polyfit} tool. \cite{polyfit}
The cQMC energies obtained for all systems can be found in the
supplementary material.

\subsection{Results}

In this section we test the \textsc{xspot} method on all eight systems
under consideration to assess the different aspects discussed in
Section \ref{sec:method} and to determine the broader applicability of
the method.
The VMC and bDMC energies and fits we obtain for the eight systems are
shown in Fig.\ \ref{fig:molecules}; additional plots containing the
bVMC and DMC energies can be found in the supplementary material.
\begin{figure*}[!hbt]
  \centering
  \includegraphics[width=0.92\columnwidth]{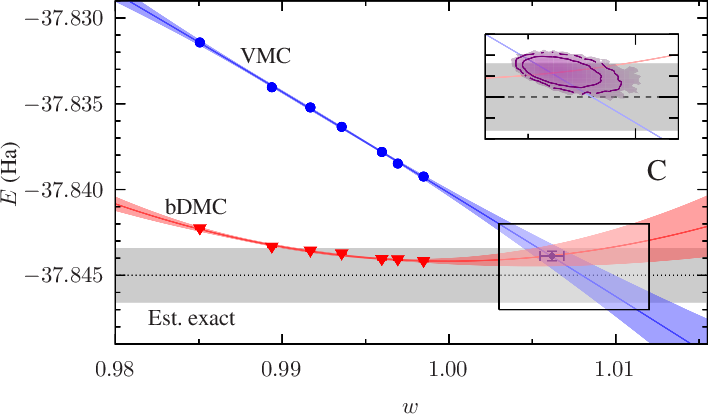} \quad
  \includegraphics[width=0.92\columnwidth]{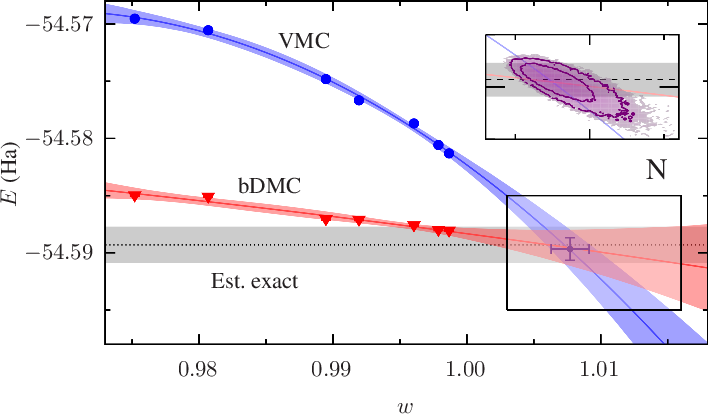} \\[0.3cm]
  \includegraphics[width=0.92\columnwidth]{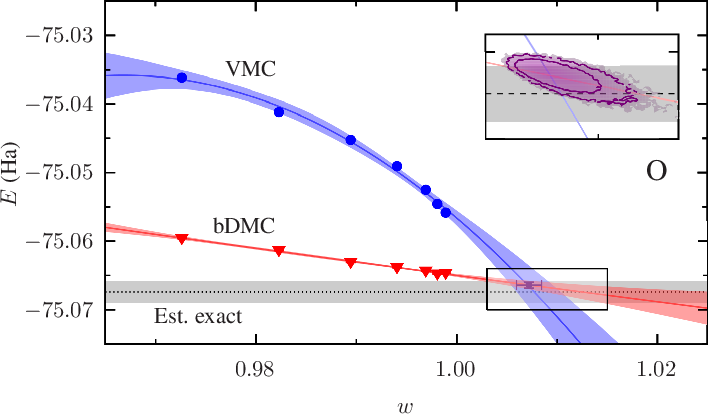} \quad
  \includegraphics[width=0.92\columnwidth]{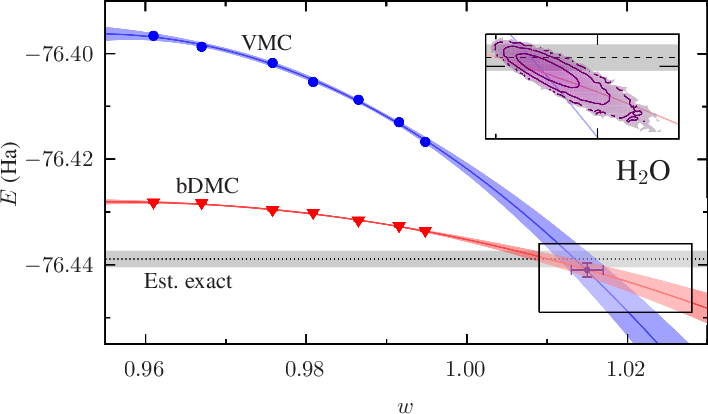} \\[0.3cm]
  \includegraphics[width=0.92\columnwidth]{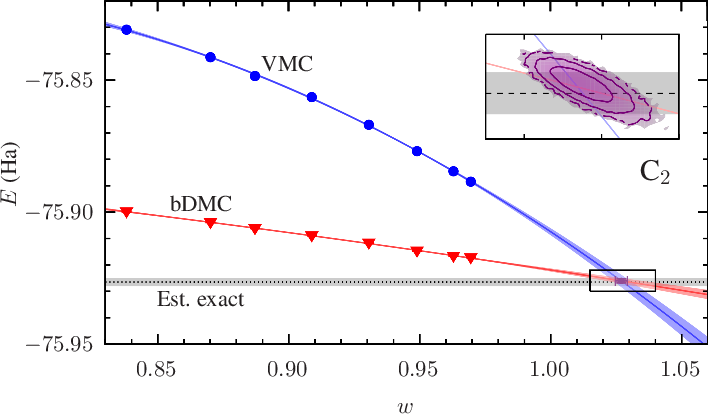} \quad
  \includegraphics[width=0.92\columnwidth]{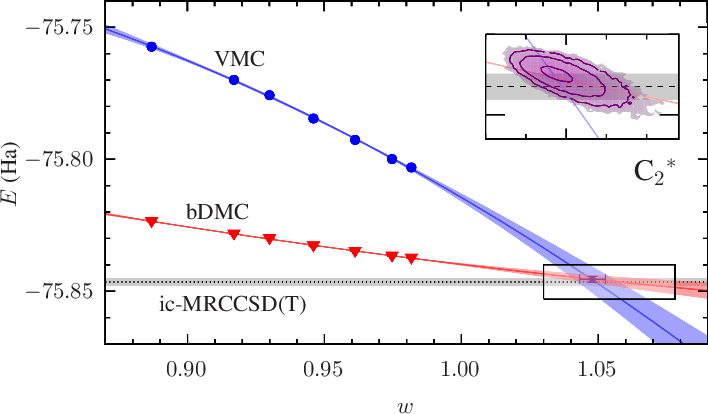} \\[0.3cm]
  \includegraphics[width=0.92\columnwidth]{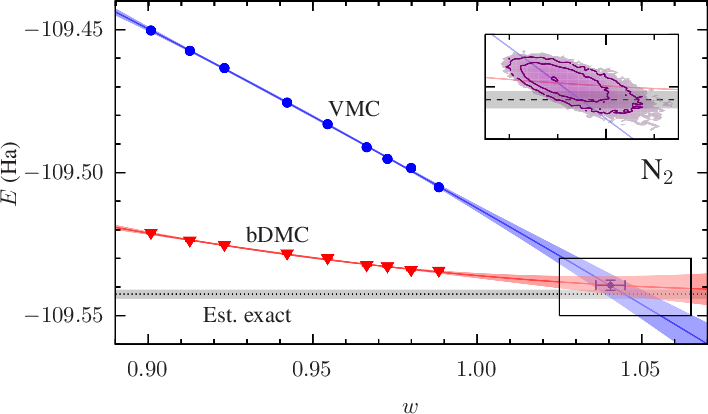} \quad
  \includegraphics[width=0.92\columnwidth]{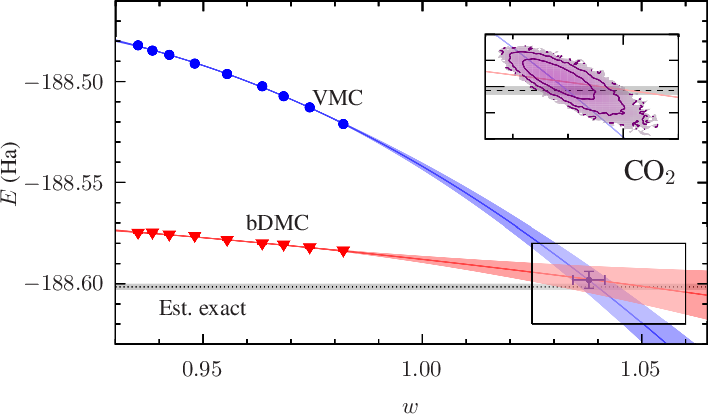}
  \caption{
    VMC and bDMC energies of the atoms and molecules considered in
    this work as a function of $w$.
    Mean values of the fits to the data are shown as lines, and the
    translucent areas around them represent 95.5\% (two-sigma)
    confidence intervals.
    Also shown in each plot is the relevant benchmark energy (see
    details in text and Table \ref{table:total_energies}) as a dotted
    line with a shaded area of $\pm 1$ kcal/mol around it, and the
    intersection point between the VMC and bDMC curves.
    The insets show the statistical distributions of intersection
    points as color maps with overlaid contour curves.
  }
  \label{fig:molecules}
\end{figure*}

All the curves in Fig.\ \ref{fig:molecules} are relatively smooth and
provide a well-defined intersection.
The apparent non-monotonicity of the bDMC curve for the carbon atom is
an artifact of the use of a fit function which formally allows
non-monotonic behavior, and should be interpreted accordingly:\@ $E(w)$
can be regarded to approach the intersection with negligible slope,
and the region $w>w_0$ should be ignored since $E(w)$ does not have a
physical meaning there.
All of the other fits appear to be monotonic in the range shown.

\begin{figure}[!hbt]
  \centering
  \includegraphics[width=\columnwidth]{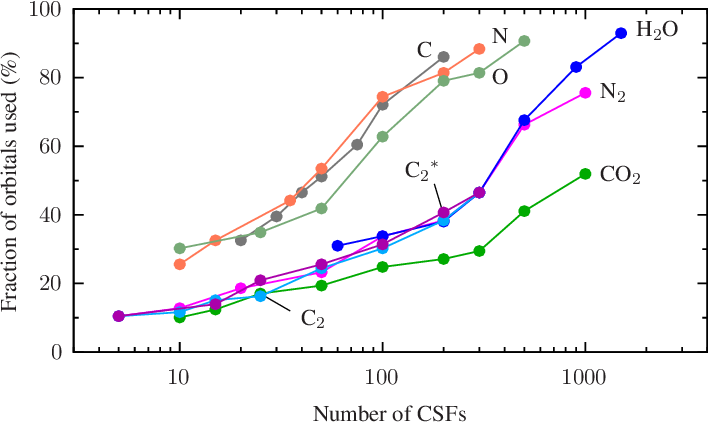}
  \caption{
    Number of distinct spatial orbitals in the FCIQMC trial wave
    functions relative to that in the orbital basis as a function of
    the number of CSFs, for the various systems considered in this
    work.
  }
  \label{fig:orb_count}
\end{figure}

The fraction of orbitals used, a proxy for the degree of orbital-basis
exhaustion, is plotted for each of the systems in Fig.\
\ref{fig:orb_count}.
We do not use up all of the orbitals in the basis in any of our
calculations, and the curves in Fig.\ \ref{fig:molecules} do not seem
to exhibit symptoms of orbital-basis exhaustion.
The bVMC energies do seem to plateau somewhat, which we discuss
briefly in the supplementary material; note that we do not use the
bVMC data to obtain our final results.

In Table \ref{table:total_energies} we compare the total energies
obtained from applying the \textsc{xspot} method to VMC and bDMC
energy data with benchmark-quality estimates of the exact
nonrelativisitic energies of the systems from the literature, along
with our best bDMC result and prior cQMC results for reference.
\begin{table*}[ht!]
  \begin{tabular}{cr@{.}lr@{.}lr@{$\phantom{.}\pm\phantom{.}$}lr@{.}l}
    \multicolumn{1}{c}{System       } &
    \multicolumn{2}{c}{Prior cQMC   } &
    \multicolumn{2}{c}{Our best bDMC} &
    \multicolumn{2}{c}{\textsc{xspot}} &
    \multicolumn{2}{c}{Benchmark    } \\
    \hline
    \hline
    C      & $ -37$&$84438(5) $ $^{\it a}$
           & $ -37$&$8442(0)$
           & $ -37.8439 $&$ 0.0003$
           & $ -37$&$8450     $ $^{\it b}$ \\
    N      & $ -54$&$58829(7) $ $^{\it a}$
           & $ -54$&$5881(0)$
           & $ -54.5897 $&$ 0.0014$
           & $ -54$&$5893     $ $^{\it b}$ \\
    O      & $ -75$&$06591(8) $ $^{\it a}$
           & $ -75$&$0647(1)$
           & $ -75.0664 $&$ 0.0004$
           & $ -75$&$0674     $ $^{\it b}$ \\
    H$_2$O & $ -76$&$4389(1)  $ $^{\it d}$
           & $ -76$&$4336(1)$
           & $ -76.4410 $&$ 0.0013$
           & $ -76$&$4389     $ $^{\it c}$ \\
    C$_2$  & $ -75$&$9229(6)  $ $^{\it a}$
           & $ -75$&$9172(1)$
           & $ -75.9260 $&$ 0.0008$
           & $ -75$&$9265     $ $^{\it b}$ \\
    ${{\textrm{C}}_2}^*$&
             \multicolumn{2}{c}{~}
           & $ -75$&$8374(1)$
           & $ -75.8455 $&$ 0.0011$
           & $ -75$&$8465     $ $^{\it e, b}$ \\
    N$_2$  & $-109$&$5372(3)  $ $^{\it a}$
           & $-109$&$5344(1)$
           & $-109.5394 $&$ 0.0018$
           & $-109$&$5425     $ $^{\it c}$ \\
    CO$_2$ & \multicolumn{2}c{~}
           & $-188$&$5837(2)$
           & $-188.5981 $&$ 0.0042$
           & $-188$&$6015     $ $^{\it c}$ \\
    \hline
    \multicolumn{7}{c}{
      $^{\it a}$Ref. \onlinecite{Morales_mdet_2012};
      $^{\it b}$Ref.\ \onlinecite{Bytautas_diatomic_2005};
      $^{\it c}$Ref.\ \onlinecite{Feller_molecules_2008};
      $^{\it d}$Ref. \onlinecite{Caffarel_h2o_2016};
      $^{\it e}$Refs.\ \onlinecite{Samanta_c2_ex_run,
                Samanta_icmrcc_ex_2014, Hanauer_icmrcc_2011}.
    }
  \end{tabular}
  \caption{
    Total energies in Ha obtained with the \textsc{xspot} method using
    the VMC and bDMC data for the various atoms and molecules
    considered in this work, along with results from prior
    multi-determinant cQMC studies, our best individual bDMC energy
    for each system, and benchmark-quality nonrelativistic total
    energies from the literature.
  }
  \label{table:total_energies}
\end{table*}
The atomization energies of the ground-state molecules are shown in
Table \ref{table:atomization_energies}.
\begin{table}[ht!]
  \begin{tabular}{r@{$\phantom{.}\to\phantom{.}$}l
                  r@{$\phantom{.}\pm\phantom{.}$}rr@{.}l}
    \multicolumn{2}{c}{                         } &
    \multicolumn{2}{c}{\textsc{xspot}           } &
    \multicolumn{2}{c}{Benchmark                } \\
    \hline
    \hline
    H$_2$O & $2$H + O
      & $374.6 $&$ 1.4 $ & $371$&$4$ $^{\it b}$ \\
    C$_2$ & $2$C
      & $238.3 $&$ 0.9 $ &~~~ $236$&$5$ $^{\it a}$ \\
    C$_2$ & ${{\textrm{C}}_2}^*$
      & $ 80.6 $&$ 1.3 $ & $ 80$&$0$ $^{\it c}$ \\
    N$_2$ & $2$N
      & $360.1 $&$ 2.3 $ & $363$&$9$ $^{\it b}$ \\
    CO$_2$ & C + $2$O
      & $621.4 $&$ 4.3 $ & $621$&$7$ $^{\it b}$ \\
    \hline
    \multicolumn{6}{c}{
      $^{\it a}$Ref.\ \onlinecite{Bytautas_diatomic_2005};
      $^{\it b}$Ref.\ \onlinecite{Feller_molecules_2008};
      $^{\it c}$Refs.\ \onlinecite{Samanta_c2_ex_run,
                Samanta_icmrcc_ex_2014, Hanauer_icmrcc_2011}.
    }
  \end{tabular}
  \caption{
    Atomization and excitation energies in mHa of the various
    molecules considered in this work corresponding to the total
    energies in Table \ref{table:total_energies} obtained from the
    \textsc{xspot} method, along with benchmark-quality
    nonrelativistic relative energies from the literature.
  }
  \label{table:atomization_energies}
\end{table}
For excited-state ${{\textrm{C}}_2}^*$ we compare the vertical
excitation energy with that calculated with internally-contracted
multi-reference coupled cluster theory (ic-MRCC);
\cite{Samanta_c2_ex_run, Samanta_icmrcc_ex_2014, Hanauer_icmrcc_2011}
we have computed the total energy of ${{\textrm{C}}_2}^*$ shown in
Table \ref{table:total_energies} by adding the ic-MRCC excitation
energy to the estimated ground-state energy of C$_2$ from Ref.\
\onlinecite{Bytautas_diatomic_2005}.

All of the total energies reported in Table \ref{table:total_energies}
are within statistical uncertainty of their corresponding benchmark
values.
An important observation is that our individual cQMC energies are not
lower than those from prior cQMC calculations, implying that we incur
a lower computational cost, but our \textsc{xspot} results
\textit{are} in general closer to the benchmarks than cQMC results
from prior studies.
Our \textsc{xspot} energies are on average $1.1$ standard errors above
the benchmark, with a root-mean-square deviation of $1.9$ standard
errors.
These results are compatible with the \textsc{xspot} method being
exact when the method's assumptions are satisfied.
The relative energies are likewise in agreement with the benchmark
values.

We find that the magnitude $\alpha$ of the quasi-random fluctuations
of the cQMC energies is of up to $0.7$ mHa.
These fluctuations are particularly visible in the VMC data for N, O,
and H$_2$O in Fig.\ \ref{fig:molecules}, for example; in the
supplementary material we give the values of $\alpha$ we have obtained
for each of the curves.
The magnitude of the quasirandom fluctuations does not seem to
increase too rapidly with system size, but their effect on the
extrapolated energy becomes more pronounced the further the cQMC data
are from the intersection in the plots.
This increasing uncertainty on the \textsc{xspot} total energies,
reaching $4$ mHa for the CO$_2$ molecule, hints at a limitation of the
methodology:\@ the cQMC energies and values of $w$ obtained using
modest-sized multideterminant expansions with a fixed basis set move
away from the intersection point with increasing system size, which in
turn exacerbates the effects of quasirandom noise on the uncertainty
of the \textsc{xspot} energy; one would have to use bigger basis sets
and larger multideterminantal expansions to get data closer to the
intersection in order to reduce this uncertainty, increasing the
computational cost of the approach.

\section{Conclusions}
\label{sec:conclusions}

We have presented an empirical extrapolation strategy for cQMC
energies as a function of the sum of the squared multideterminant
coefficients in the initial selected CI wave function from which the
trial wave function is constructed.
This approach is made possible by the smoothness of the energies as a
function of the CI expansion size, and we have presented a simple
statistical procedure to handle the quasi-random non-smoothness in the
data, which we show to work very well in practice.
We find that Hartree-Fock orbitals expanded in standard basis sets
provide the type of gradual convergence required for the
\textsc{xspot} method to work well.
The results from the tests we have conducted are compatible with the
\textsc{xspot} method being capable of obtaining exact total energies,
with the caveat that trial wave function complexity must increase with
system size in order to control the uncertainty on the results.

\section*{Supplementary Material}

See the supplementary material for the cQMC data used in this paper, a
table of the magnitude of the quasirandom fluctuations encountered,
and discussion of connected extrapolation approaches.
The supplementary material additionally cites Refs.\
\onlinecite{Holmes_shci_extrap_2017, Burton_cipsi_extrap_2024,
Ceperley_1986, Taddei_2015, Fu_2024}.

\begin{acknowledgments}
P.L.R. and A.A. acknowledge support from the European Centre of
Excellence in Exascale Computing TREX, funded by the Horizon 2020
program of the European Union under grant no.\ 952165.
Views and opinions expressed are those of the authors only and do not
necessarily reflect those of the European Union or the European
Research Executive Agency.
Neither the European Union nor the granting authority can be held
responsible for them.
\end{acknowledgments}


\setcounter{table}{0}
\renewcommand{\thetable}{S\arabic{table}}
\setcounter{figure}{0}
\renewcommand{\thefigure}{S\arabic{figure}}
\setcounter{section}{0}
\renewcommand{\thesection}{S\arabic{section}}
\setcounter{equation}{0}
\renewcommand{\theequation}{S\arabic{equation}}
\renewcommand{\bibnumfmt}[1]{[S#1]}
\renewcommand{\citenumfont}[1]{S#1}

\onecolumngrid

\pagebreak
\begin{center}
\textbf{\large Supplemental information for ``X marks the spot:\@
   accurate energies from intersecting extrapolations of continuum
   quantum Monte Carlo data''}
\end{center}

\section{VMC and DMC data}

\begin{figure}[!hbt]
  \centering
  \includegraphics[width=0.46\columnwidth]{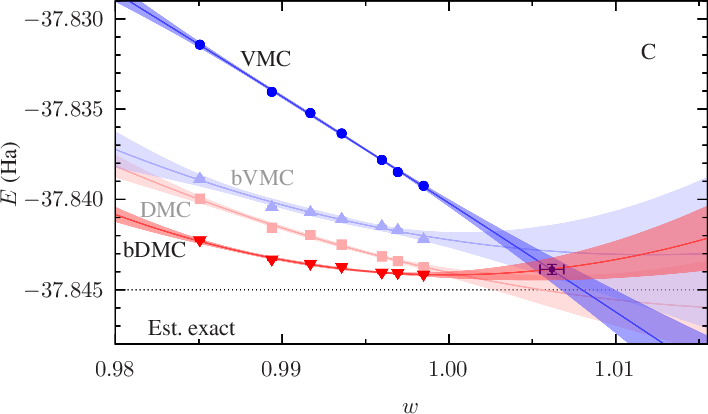} \quad
  \includegraphics[width=0.46\columnwidth]{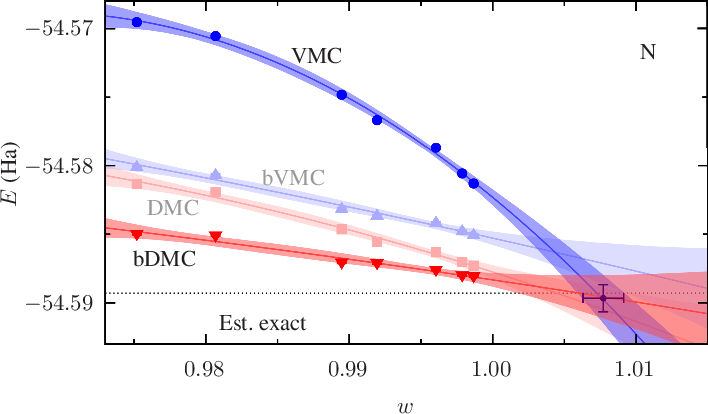} \\[0.3cm]
  \includegraphics[width=0.46\columnwidth]{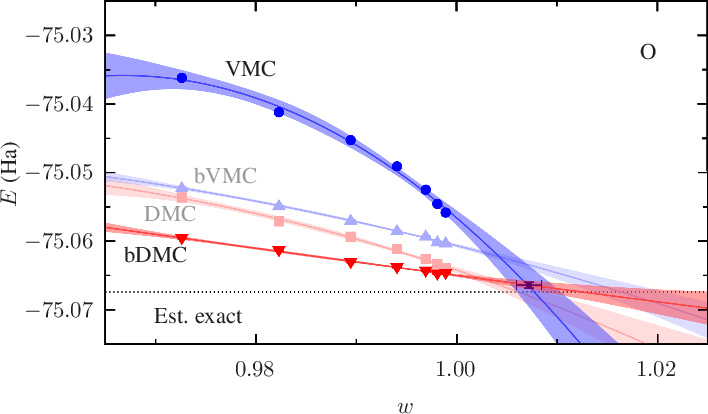} \quad
  \includegraphics[width=0.46\columnwidth]{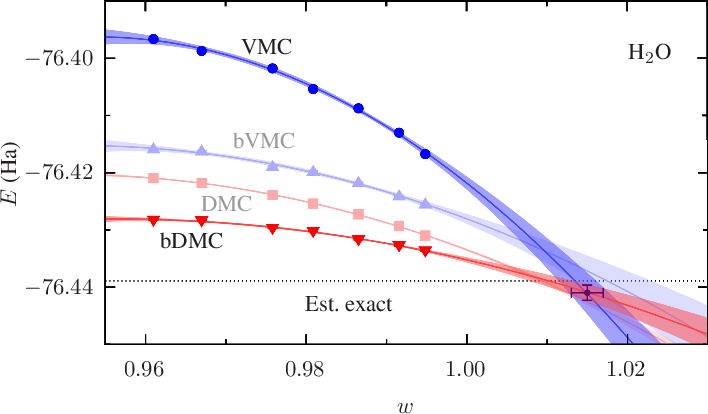} \\[0.3cm]
  \includegraphics[width=0.46\columnwidth]{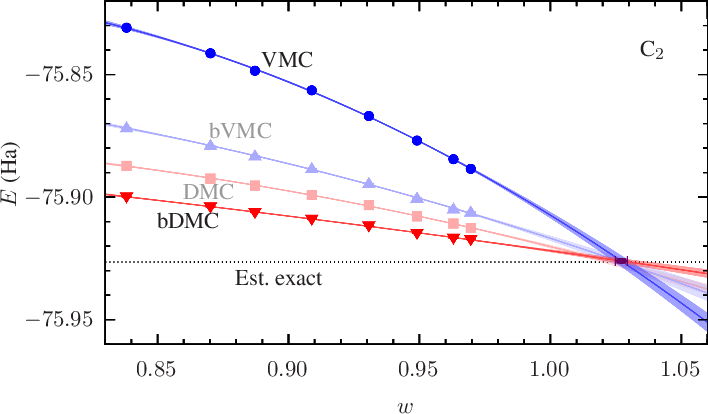} \quad
  \includegraphics[width=0.46\columnwidth]{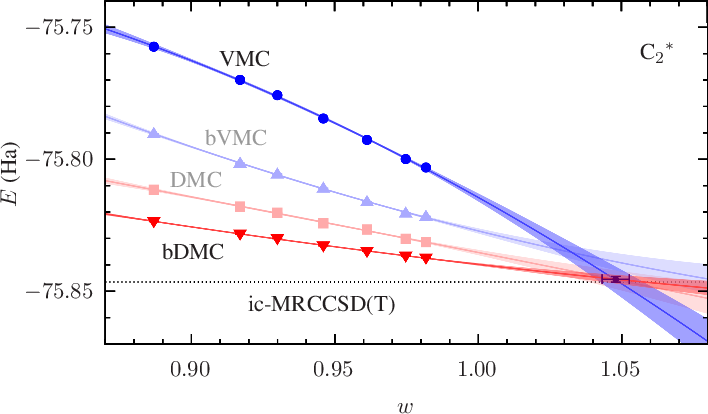} \\[0.3cm]
  \includegraphics[width=0.46\columnwidth]{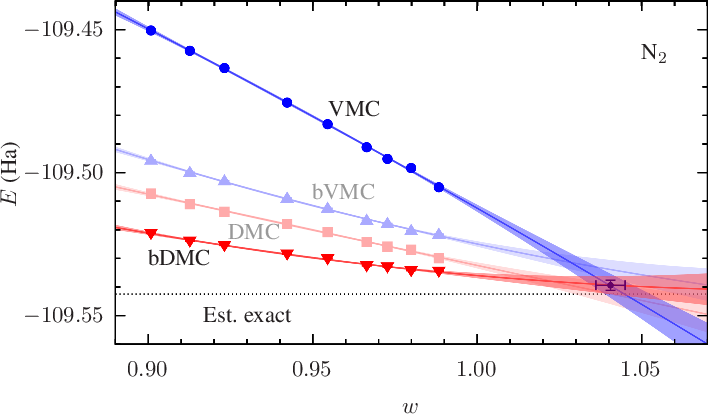} \quad
  \includegraphics[width=0.46\columnwidth]{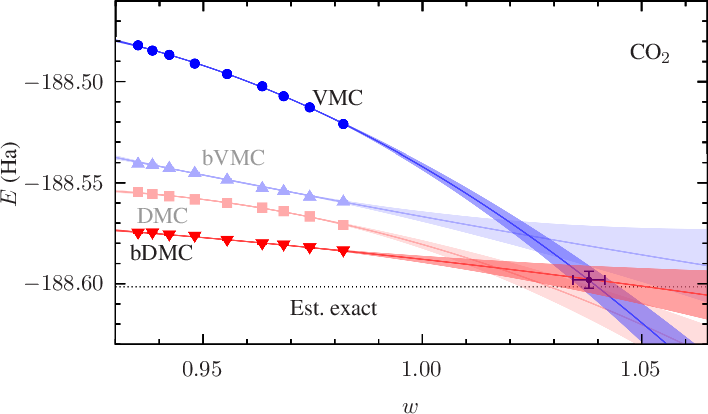} 
  \caption{ 
    VMC, DMC, bVMC, and bDMC energies of the atoms and molecules
    considered in this work as a function of $w$. 
    Mean values of the fits to the data are shown as lines, and the
    translucent areas around them represent 95.5\% (two-sigma)
    confidence intervals.
    Also shown in each plot is the relevant benchmark energy (see
    details in text and Table 4 of the manuscript) as a dotted
    line with a shaded area of $\pm 1$ kcal/mol around it, and the
    intersection point between the VMC and bDMC curves.
  }   
  \label{fig:molecules_4curve}
\end{figure}

Tables \ref{table:c_atom}--\ref{table:co2} contain the cQMC energies
we have obtained for our manuscript.
These energies are plotted in Fig.\ \ref{fig:molecules_4curve}, as
Fig.\ 4 of the manuscript with added backflow VMC and non-backflow DMC
curves.

The bVMC energy curves in Fig.\ \ref{fig:molecules_4curve} would
appear to plateau somewhat at large $w$, a potential symptom of
orbital-basis exhaustion.
We hypothesize that the backflow transformation makes some of the
information contained in the orbitals redundant, which effectively
reduces the value of $w_{\rm exh.}$ with respect to the non-backflow
data.
Judging by the quality of the intersections, this issue affects bVMC
more than it does bDMC, which would imply that this redundancy does
not involve the location of the nodes of the trial wave function to
the same degree as its values away from the nodes. 
This is in any case a tentative explanation; we do not use the bVMC
data to obtain our final results.

\begin{table}[ht!]
  \begin{tabular}{rrrrr@{.}lr@{.}lr@{.}lr@{.}l}
    \multicolumn{1}{c}{$n_{\rm CSF}$} &
    \multicolumn{1}{c}{$n_{\rm det}$} &
    \multicolumn{1}{c}{$n_{\rm orb}$} &
    \multicolumn{1}{c}{$w$          } &
    \multicolumn{2}{c}{VMC}           &
    \multicolumn{2}{c}{DMC}           &
    \multicolumn{2}{c}{bVMC}          &
    \multicolumn{2}{c}{bDMC}          \\
    \hline
    \hline
     20 &  45 & 14 & $0.9851$ & $-37$&$83143(2)$ & $-37$&$83996(6)$ & $-37$&$83884(2)$ & $-37$&$84228(3)$ \\
     30 &  68 & 17 & $0.9894$ & $-37$&$83404(2)$ & $-37$&$84157(6)$ & $-37$&$84039(2)$ & $-37$&$84335(3)$ \\
     40 &  87 & 20 & $0.9917$ & $-37$&$83522(2)$ & $-37$&$84197(5)$ & $-37$&$84068(2)$ & $-37$&$84358(3)$ \\
     50 & 109 & 22 & $0.9936$ & $-37$&$83635(2)$ & $-37$&$84250(5)$ & $-37$&$84106(2)$ & $-37$&$84374(3)$ \\
     75 & 155 & 26 & $0.9960$ & $-37$&$83781(2)$ & $-37$&$84316(4)$ & $-37$&$84146(2)$ & $-37$&$84407(3)$ \\
    100 & 222 & 31 & $0.9969$ & $-37$&$83849(2)$ & $-37$&$84342(4)$ & $-37$&$84167(2)$ & $-37$&$84409(3)$ \\
    200 & 476 & 37 & $0.9985$ & $-37$&$83924(2)$ & $-37$&$84375(4)$ & $-37$&$84219(2)$ & $-37$&$84418(3)$ \\
    \multicolumn{4}{c}{Benchmark} & \multicolumn{8}{c}{$-37.8450$} \\
    \hline
  \end{tabular}
  \caption{
    VMC, DMC, bVMC, and bDMC energies for the C atom obtained for our
    manuscript, in Hartree atomic units.
    $n_{\rm CSF}$ is the number of CSFs in the wave function,
    $n_{\rm det}$ is the number of (not necessarily unique)
    determinants, and $n_{\rm orb}$ is the number of orbitals
    from the 43-orbital basis used in the wave function.
  }
  \label{table:c_atom}
\end{table}

\begin{table}[ht!]
  \begin{tabular}{rrrrr@{.}lr@{.}lr@{.}lr@{.}l}
    \multicolumn{1}{c}{$n_{\rm CSF}$} &
    \multicolumn{1}{c}{$n_{\rm det}$} &
    \multicolumn{1}{c}{$n_{\rm orb}$} &
    \multicolumn{1}{c}{$w$          } &
    \multicolumn{2}{c}{VMC}           &
    \multicolumn{2}{c}{DMC}           &
    \multicolumn{2}{c}{bVMC}          &
    \multicolumn{2}{c}{bDMC}          \\
    \hline
    \hline
     10 &  19 & 11 & $0.9752$ & $-54$&$56953(7)$ & $-54$&$58134(7)$ & $-54$&$58010(6)$ & $-54$&$58501(4)$ \\
     15 &  29 & 14 & $0.9807$ & $-54$&$57054(7)$ & $-54$&$58192(7)$ & $-54$&$58067(6)$ & $-54$&$58512(4)$ \\
     35 &  77 & 19 & $0.9895$ & $-54$&$57482(7)$ & $-54$&$58464(6)$ & $-54$&$58314(6)$ & $-54$&$58708(6)$ \\
     50 &  96 & 23 & $0.9919$ & $-54$&$57667(6)$ & $-54$&$58557(6)$ & $-54$&$58363(6)$ & $-54$&$58713(5)$ \\
    100 & 189 & 32 & $0.9960$ & $-54$&$57869(6)$ & $-54$&$58632(6)$ & $-54$&$58414(6)$ & $-54$&$58762(4)$ \\
    200 & 479 & 35 & $0.9979$ & $-54$&$58057(6)$ & $-54$&$58702(5)$ & $-54$&$58477(6)$ & $-54$&$58802(4)$ \\
    300 & 761 & 38 & $0.9987$ & $-54$&$58131(7)$ & $-54$&$58729(5)$ & $-54$&$58503(6)$ & $-54$&$58808(4)$ \\
    \multicolumn{4}{c}{Benchmark} & \multicolumn{8}{c}{$-54.5893$} \\
    \hline
  \end{tabular}
  \caption{
    VMC, DMC, bVMC, and bDMC energies for the N atom obtained for our
    manuscript, in Hartree atomic units.
    $n_{\rm CSF}$ is the number of CSFs in the wave function,
    $n_{\rm det}$ is the number of (not necessarily unique)
    determinants, and $n_{\rm orb}$ is the number of orbitals
    from the 43-orbital basis used in the wave function.
  }
  \label{table:n_atom}
\end{table}

\begin{table}[ht!]
  \begin{tabular}{rrrrr@{.}lr@{.}lr@{.}lr@{.}l}
    \multicolumn{1}{c}{$n_{\rm CSF}$} &
    \multicolumn{1}{c}{$n_{\rm det}$} &
    \multicolumn{1}{c}{$n_{\rm orb}$} &
    \multicolumn{1}{c}{$w$          } &
    \multicolumn{2}{c}{VMC}           &
    \multicolumn{2}{c}{DMC}           &
    \multicolumn{2}{c}{bVMC}          &
    \multicolumn{2}{c}{bDMC}          \\
    \hline
    \hline
     10 &   20 & 13 & $0.9726$ & $-75$&$0362(1)$ & $-75$&$0536(1)$ & $-75$&$05225(9)$ & $-75$&$05959(5)$ \\
     25 &   61 & 15 & $0.9823$ & $-75$&$0412(1)$ & $-75$&$0571(1)$ & $-75$&$05488(9)$ & $-75$&$06135(4)$ \\
     50 &  121 & 18 & $0.9894$ & $-75$&$0452(1)$ & $-75$&$0594(1)$ & $-75$&$05701(9)$ & $-75$&$06307(5)$ \\
    100 &  244 & 27 & $0.9940$ & $-75$&$0491(1)$ & $-75$&$0611(1)$ & $-75$&$05851(9)$ & $-75$&$06384(5)$ \\
    200 &  538 & 34 & $0.9969$ & $-75$&$0525(1)$ & $-75$&$0626(1)$ & $-75$&$05928(9)$ & $-75$&$06434(5)$ \\
    300 &  856 & 35 & $0.9981$ & $-75$&$0546(1)$ & $-75$&$0634(1)$ & $-75$&$06013(9)$ & $-75$&$06476(5)$ \\
    500 & 1598 & 39 & $0.9989$ & $-75$&$0558(1)$ & $-75$&$0639(1)$ & $-75$&$06029(9)$ & $-75$&$06467(5)$ \\
    \multicolumn{4}{c}{Benchmark} & \multicolumn{8}{c}{$-75.0674$} \\
    \hline
  \end{tabular}
  \caption{
    VMC, DMC, bVMC, and bDMC energies for the O atom obtained for our
    manuscript, in Hartree atomic units.
    $n_{\rm CSF}$ is the number of CSFs in the wave function,
    $n_{\rm det}$ is the number of (not necessarily unique)
    determinants, and $n_{\rm orb}$ is the number of orbitals
    from the 43-orbital basis used in the wave function.
  }
  \label{table:o_atom}
\end{table}

\begin{table}[ht!]
  \begin{tabular}{rrrrr@{.}lr@{.}lr@{.}lr@{.}l}
    \multicolumn{1}{c}{$n_{\rm CSF}$} &
    \multicolumn{1}{c}{$n_{\rm det}$} &
    \multicolumn{1}{c}{$n_{\rm orb}$} &
    \multicolumn{1}{c}{$w$          } &
    \multicolumn{2}{c}{VMC}           &
    \multicolumn{2}{c}{DMC}           &
    \multicolumn{2}{c}{bVMC}          &
    \multicolumn{2}{c}{bDMC}          \\
    \hline
    \hline
      60 &  190 & 22 & $0.9610$ & $-76$&$39663(3)$ & $-76$&$4209(8)$ & $-76$&$4159(1)$ & $-76$&$42825(7)$ \\
     100 &  338 & 24 & $0.9670$ & $-76$&$3987(1) $ & $-76$&$4218(8)$ & $-76$&$4162(1)$ & $-76$&$42836(7)$ \\
     200 &  702 & 27 & $0.9758$ & $-76$&$4018(1) $ & $-76$&$4239(8)$ & $-76$&$4190(1)$ & $-76$&$42968(7)$ \\
     300 & 1079 & 33 & $0.9809$ & $-76$&$4054(1) $ & $-76$&$4254(7)$ & $-76$&$4198(1)$ & $-76$&$43022(8)$ \\
     500 & 1798 & 48 & $0.9865$ & $-76$&$4088(1) $ & $-76$&$4273(7)$ & $-76$&$4218(1)$ & $-76$&$43167(7)$ \\
     900 & 3298 & 59 & $0.9916$ & $-76$&$4130(1) $ & $-76$&$4293(6)$ & $-76$&$4241(1)$ & $-76$&$43272(8)$ \\
    1500 & 5698 & 66 & $0.9948$ & $-76$&$4167(1) $ & $-76$&$4310(6)$ & $-76$&$4255(1)$ & $-76$&$43362(8)$ \\
    \multicolumn{4}{c}{Benchmark} & \multicolumn{8}{c}{$-76.4389$} \\
    \hline
  \end{tabular}
  \caption{
    VMC, DMC, bVMC, and bDMC energies for the H$_2$O molecule obtained
    for our manuscript, in Hartree atomic units.
    $n_{\rm CSF}$ is the number of CSFs in the wave function,
    $n_{\rm det}$ is the number of (not necessarily unique)
    determinants, and $n_{\rm orb}$ is the number of orbitals
    from the 71-orbital basis used in the wave function.
  }
  \label{table:h2o}
\end{table}

\begin{table}[ht!]
  \begin{tabular}{rrrrr@{.}lr@{.}lr@{.}lr@{.}l}
    \multicolumn{1}{c}{$n_{\rm CSF}$} &
    \multicolumn{1}{c}{$n_{\rm det}$} &
    \multicolumn{1}{c}{$n_{\rm orb}$} &
    \multicolumn{1}{c}{$w$          } &
    \multicolumn{2}{c}{VMC}           &
    \multicolumn{2}{c}{DMC}           &
    \multicolumn{2}{c}{bVMC}          &
    \multicolumn{2}{c}{bDMC}          \\
    \hline
    \hline
      5 &   16 &  9 & $0.8381$ & $-75$&$83089(7)$ & $-75$&$8872(1)$ & $-75$&$8718(2) $ & $-75$&$8998(1) $ \\
     10 &   28 & 10 & $0.8701$ & $-75$&$84133(7)$ & $-75$&$8924(1)$ & $-75$&$87903(8)$ & $-75$&$90389(9)$ \\
     15 &   44 & 13 & $0.8871$ & $-75$&$84845(7)$ & $-75$&$8953(1)$ & $-75$&$88329(7)$ & $-75$&$90615(9)$ \\
     25 &   68 & 14 & $0.9088$ & $-75$&$85644(6)$ & $-75$&$8991(1)$ & $-75$&$8885(1) $ & $-75$&$90891(9)$ \\
     50 &  162 & 21 & $0.9306$ & $-75$&$86691(6)$ & $-75$&$9033(1)$ & $-75$&$8946(1) $ & $-75$&$91166(8)$ \\
    100 &  350 & 26 & $0.9490$ & $-75$&$87693(6)$ & $-75$&$9079(1)$ & $-75$&$9005(1) $ & $-75$&$91468(7)$ \\
    200 &  730 & 33 & $0.9629$ & $-75$&$88454(6)$ & $-75$&$9108(1)$ & $-75$&$9049(1) $ & $-75$&$91665(7)$ \\
    300 & 1160 & 40 & $0.9695$ & $-75$&$88847(6)$ & $-75$&$9126(1)$ & $-75$&$9062(1) $ & $-75$&$91724(7)$ \\
    \multicolumn{4}{c}{Benchmark} & \multicolumn{8}{c}{$-75.9265$} \\
    \hline
  \end{tabular}
  \caption{
    VMC, DMC, bVMC, and bDMC energies for the ground-state C$_2$
    molecule obtained for our manuscript, in Hartree atomic units.
    $n_{\rm CSF}$ is the number of CSFs in the wave function,
    $n_{\rm det}$ is the number of (not necessarily unique)
    determinants, and $n_{\rm orb}$ is the number of orbitals
    from the 86-orbital basis used in the wave function.
  }
  \label{table:c2}
\end{table}

\begin{table}[ht!]
  \begin{tabular}{rrrrr@{.}lr@{.}lr@{.}lr@{.}l}
    \multicolumn{1}{c}{$n_{\rm CSF}$} &
    \multicolumn{1}{c}{$n_{\rm det}$} &
    \multicolumn{1}{c}{$n_{\rm orb}$} &
    \multicolumn{1}{c}{$w$          } &
    \multicolumn{2}{c}{VMC}           &
    \multicolumn{2}{c}{DMC}           &
    \multicolumn{2}{c}{bVMC}          &
    \multicolumn{2}{c}{bDMC}          \\
    \hline
    \hline
      5 &    7 &  9 & $0.8869$ & $-75$&$75737(8)$ & $-75$&$8116(1) $ & $-75$&$7904(1) $ & $-75$&$82356(8)$ \\
     15 &   37 & 12 & $0.9170$ & $-75$&$76993(8)$ & $-75$&$8179(1) $ & $-75$&$8016(1) $ & $-75$&$82828(9)$ \\
     25 &   75 & 18 & $0.9299$ & $-75$&$77575(8)$ & $-75$&$8203(1) $ & $-75$&$80570(9)$ & $-75$&$83000(8)$ \\
     50 &  167 & 22 & $0.9460$ & $-75$&$78458(8)$ & $-75$&$8243(1) $ & $-75$&$81104(9)$ & $-75$&$83274(8)$ \\
    100 &  348 & 27 & $0.9612$ & $-75$&$79267(8)$ & $-75$&$82662(9)$ & $-75$&$81616(9)$ & $-75$&$8348(1) $ \\
    200 &  690 & 35 & $0.9747$ & $-75$&$79994(8)$ & $-75$&$83013(8)$ & $-75$&$82048(8)$ & $-75$&$83661(9)$ \\
    300 & 1080 & 40 & $0.9817$ & $-75$&$80313(7)$ & $-75$&$83142(8)$ & $-75$&$82172(8)$ & $-75$&$8374(1) $ \\
    \multicolumn{4}{c}{Benchmark} & \multicolumn{8}{c}{$-75.8465$} \\
    \hline
  \end{tabular}
  \caption{
    VMC, DMC, bVMC, and bDMC energies for the excited-state
    ${{\textrm{C}}_2}^*$ molecule obtained for our manuscript, in
    Hartree atomic units.
    $n_{\rm CSF}$ is the number of CSFs in the wave function,
    $n_{\rm det}$ is the number of (not necessarily unique)
    determinants, and $n_{\rm orb}$ is the number of orbitals
    from the 86-orbital basis used in the wave function.
  }
  \label{table:c2_ex}
\end{table}

\begin{table}[ht!]
  \begin{tabular}{rrrrr@{.}lr@{.}lr@{.}lr@{.}l}
    \multicolumn{1}{c}{$n_{\rm CSF}$} &
    \multicolumn{1}{c}{$n_{\rm det}$} &
    \multicolumn{1}{c}{$n_{\rm orb}$} &
    \multicolumn{1}{c}{$w$          } &
    \multicolumn{2}{c}{VMC}           &
    \multicolumn{2}{c}{DMC}           &
    \multicolumn{2}{c}{bVMC}          &
    \multicolumn{2}{c}{bDMC}          \\
    \hline
    \hline
       5 &   13 &  9 & $0.9008$ & $-109$&$4503(1)$ & $-109$&$5073(2)$ & $-109$&$4957(1)$ & $-109$&$52120(9)$ \\
      10 &   23 & 11 & $0.9126$ & $-109$&$4574(1)$ & $-109$&$5111(1)$ & $-109$&$5000(1)$ & $-109$&$5239(1) $ \\
      20 &   71 & 16 & $0.9231$ & $-109$&$4634(1)$ & $-109$&$5137(2)$ & $-109$&$5029(1)$ & $-109$&$5254(1) $ \\
      50 &  177 & 20 & $0.9422$ & $-109$&$4755(1)$ & $-109$&$5180(1)$ & $-109$&$5090(1)$ & $-109$&$5284(1) $ \\
     100 &  355 & 29 & $0.9545$ & $-109$&$4831(1)$ & $-109$&$5208(1)$ & $-109$&$5126(1)$ & $-109$&$5300(1) $ \\
     200 &  748 & 33 & $0.9664$ & $-109$&$4911(1)$ & $-109$&$5243(1)$ & $-109$&$5167(1)$ & $-109$&$5325(1) $ \\
     300 & 1145 & 40 & $0.9727$ & $-109$&$4952(1)$ & $-109$&$5259(2)$ & $-109$&$5179(1)$ & $-109$&$5329(1) $ \\
     500 & 1199 & 57 & $0.9799$ & $-109$&$4984(1)$ & $-109$&$5270(2)$ & $-109$&$5203(1)$ & $-109$&$5342(1) $ \\
    1000 & 4148 & 65 & $0.9883$ & $-109$&$5051(1)$ & $-109$&$5299(2)$ & $-109$&$5216(1)$ & $-109$&$5344(1) $ \\
    \multicolumn{4}{c}{Benchmark} & \multicolumn{8}{c}{$-109.5425$} \\
    \hline
  \end{tabular}
  \caption{
    VMC, DMC, bVMC, and bDMC energies for the N$_2$ molecule obtained
    for our manuscript, in Hartree atomic units.
    $n_{\rm CSF}$ is the number of CSFs in the wave function,
    $n_{\rm det}$ is the number of (not necessarily unique)
    determinants, and $n_{\rm orb}$ is the number of orbitals
    from the 86-orbital basis used in the wave function.
  }
  \label{table:n2}
\end{table}

\begin{table}[ht!]
  \begin{tabular}{rrrrr@{.}lr@{.}lr@{.}lr@{.}l}
    \multicolumn{1}{c}{$n_{\rm CSF}$} &
    \multicolumn{1}{c}{$n_{\rm det}$} &
    \multicolumn{1}{c}{$n_{\rm orb}$} &
    \multicolumn{1}{c}{$w$          } &
    \multicolumn{2}{c}{VMC}           &
    \multicolumn{2}{c}{DMC}           &
    \multicolumn{2}{c}{bVMC}          &
    \multicolumn{2}{c}{bDMC}          \\
    \hline
    \hline
      10 &   23 & 13 & $0.9352$ & $-188$&$48208(8)$ & $-188$&$5546(2)$ & $-188$&$54035(8)$ & $-188$&$5748(1)$ \\
      15 &   43 & 16 & $0.9384$ & $-188$&$48468(8)$ & $-188$&$5556(2)$ & $-188$&$54091(8)$ & $-188$&$5748(1)$ \\
      25 &   85 & 22 & $0.9423$ & $-188$&$48678(8)$ & $-188$&$5567(2)$ & $-188$&$54247(8)$ & $-188$&$5758(1)$ \\
      50 &  191 & 25 & $0.9481$ & $-188$&$49112(8)$ & $-188$&$5582(2)$ & $-188$&$54495(8)$ & $-188$&$5764(1)$ \\
     100 &  390 & 32 & $0.9555$ & $-188$&$49628(8)$ & $-188$&$5600(2)$ & $-188$&$54827(8)$ & $-188$&$5784(1)$ \\
     200 &  820 & 35 & $0.9635$ & $-188$&$50237(8)$ & $-188$&$5623(2)$ & $-188$&$55245(9)$ & $-188$&$5802(2)$ \\
     300 & 1205 & 38 & $0.9684$ & $-188$&$50730(8)$ & $-188$&$5641(2)$ & $-188$&$55395(9)$ & $-188$&$5807(2)$ \\
     500 & 2016 & 53 & $0.9743$ & $-188$&$51281(8)$ & $-188$&$5667(2)$ & $-188$&$5567(1) $ & $-188$&$5821(2)$ \\
    1000 & 4177 & 67 & $0.9820$ & $-188$&$52106(9)$ & $-188$&$5709(1)$ & $-188$&$5591(1) $ & $-188$&$5837(2)$ \\
    \multicolumn{4}{c}{Benchmark} & \multicolumn{8}{c}{$-188.6015$} \\
    \hline
  \end{tabular}
  \caption{
    VMC, DMC, bVMC, and bDMC energies for the CO$_2$ molecule obtained
    for our manuscript, in Hartree atomic units.
    $n_{\rm CSF}$ is the number of CSFs in the wave function,
    $n_{\rm det}$ is the number of (not necessarily unique)
    determinants, and $n_{\rm orb}$ is the number of orbitals
    from the 129-orbital basis used in the wave function.
  }
  \label{table:co2}
\end{table}

\FloatBarrier

\FloatBarrier
\section{Values of $\alpha$}

Table \ref{table:alpha} shows the magnitude of the quasirandom
fluctuations $\alpha$ estimated for each of the curves we have
obtained, as defined in Eq.\ 7 of the manuscript.
Note that all the individual cQMC calculations reported in Tables
\ref{table:c_atom}--\ref{table:co2} have been run for long enough that
the statistical uncertainties on them are smaller then the
corresponding value of $\alpha$.
\begin{table}[ht!]
  \begin{tabular}{ccccc}
    System & VMC     & DMC     & bVMC    & bDMC    \\
    \hline\hline
    C      & $0.087$ & $0.096$ & $0.181$ & $0.069$ \\
    N      & $0.330$ & $0.288$ & $0.275$ & $0.283$ \\
    O      & $0.708$ & $0.265$ & $0.132$ & $0.136$ \\
    H$_2$O & $0.408$ & $0.020$ & $0.302$ & $0.144$ \\
    C$_2$  & $0.410$ & $0.193$ & $0.321$ & $0.170$ \\
    ${{\textrm{C}}_2}^*$
           & $0.508$ & $0.339$ & $0.325$ & $0.136$ \\
    N$_2$  & $0.492$ & $0.377$ & $0.367$ & $0.344$ \\
    CO$_2$ & $0.337$ & $0.360$ & $0.470$ & $0.271$ \\
    \hline
  \end{tabular}
  \caption{
    Magnitude of the quasirandom error $\alpha$ for the VMC, DMC,
    bVMC, and bDMC curves used in our manuscript, in mHa.
  }
  \label{table:alpha}
\end{table}

\FloatBarrier
\section{Connection with selected CI and DMC variance extrapolation}

In selected CI methods one usually obtains the full-CI limit by
extrapolates the total energy to the limit where the difference
between the total energy and the variational energy (which typically
amounts to a perturbation-theory correction) is zero.
\cite{Holmes_shci_extrap_2017, Burton_cipsi_extrap_2024}
In the same spirit, one could consider extrapolating the DMC energy to
the limit where the VMC-DMC energy difference is zero, which, by
analogy with the selected CI approach, would effectively treat DMC as
a perturbative correction of sorts on top of VMC.

Note that $E_{\rm VMC}-E_{\rm DMC}$ is proportional to the DMC
variance, \cite{Ceperley_1986} so extrapolating to $E_{\rm VMC}-E_{\rm
DMC}\to 0$ amounts to DMC variance extrapolation, analogous to VMC
variance extrapolation schemes which have been used over the years,
recent examples of which include Refs.\ \onlinecite{Taddei_2015,
Fu_2024}.

One potential advantage of the DMC variance extrapolation approach
over \textsc{xspot} is that the ``exact'' limit corresponds to a
predefined value (zero) of the independent variable $E_{\rm
VMC}-E_{\rm DMC}$, instead of an unknown value $w_0$, so one does not
need to explicitly intersect curves.

Clearly $E_{\rm VMC}-E_{\rm DMC}=0$ implies that the VMC and DMC
energy curves intersect as a function of $w$.
From \textsc{xspot}, one can in fact derive that the DMC energy can be
approximated by a second-order polynomial in $E_{\rm VMC}-E_{\rm
DMC}$, but since $w$ is eliminated in this process, the fit form for
DMC variance extrapolation expresses no explicit assumptions on the
dependence of the cQMC energies on the wave function.
On the one hand, this is conceptually simpler, and could in principle
avoid the orbital-basis exhaustion issue of the \textsc{xspot} method,
since if both the VMC and DMC curves exhibit plateaus as functions of
$w$ then the DMC variance extrapolation curve may stall but need not
deform.
On the other hand, the explicit assumptions made by \textsc{xspot}
could result in useful restrictions that guide the fits and reduce the
uncertainty on the final results, an improvement that DMC variance
extrapolation would miss on.

\begin{figure}[!hbt]
  \centering
  \includegraphics[width=0.46\columnwidth]{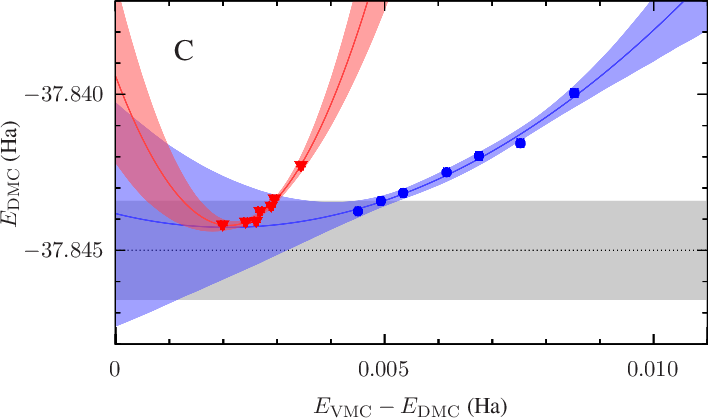} \quad
  \includegraphics[width=0.46\columnwidth]{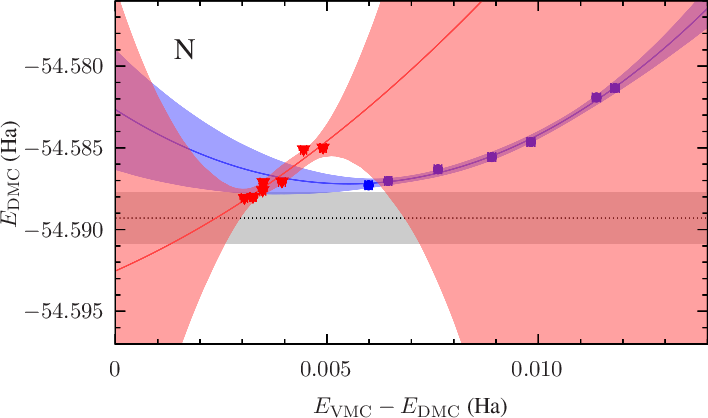} \\[0.3cm]
  \includegraphics[width=0.46\columnwidth]{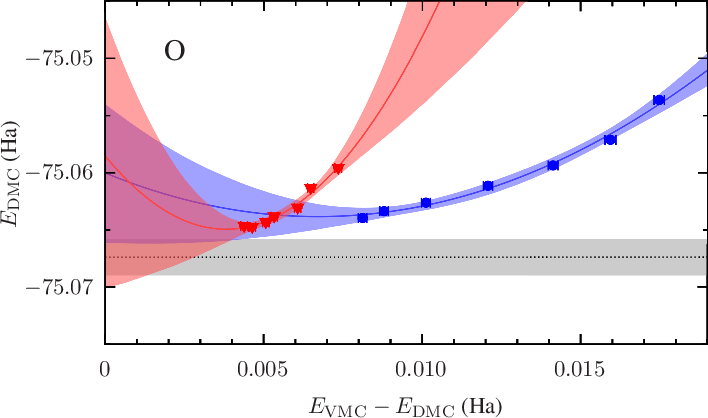} \quad
  \includegraphics[width=0.46\columnwidth]{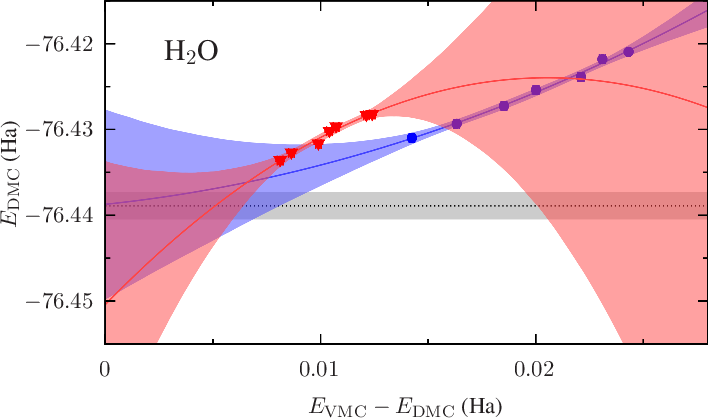} \\[0.3cm]
  \includegraphics[width=0.46\columnwidth]{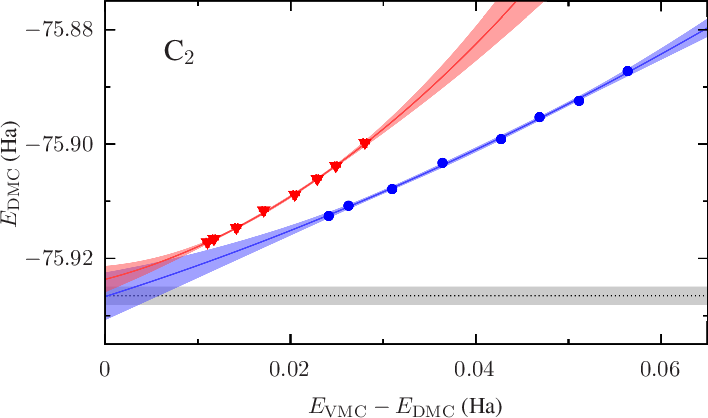} \quad
  \includegraphics[width=0.46\columnwidth]{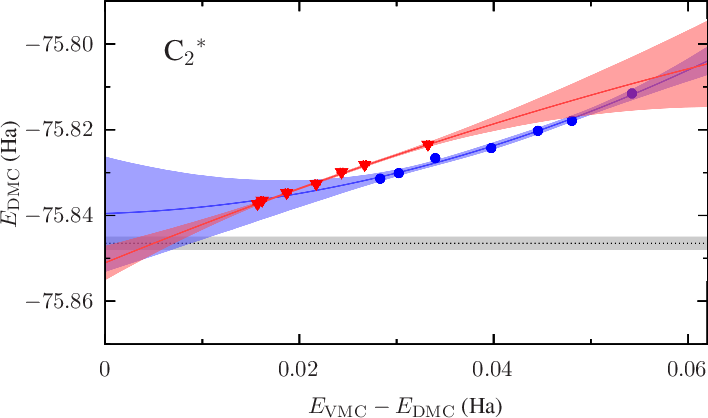} \\[0.3cm]
  \includegraphics[width=0.46\columnwidth]{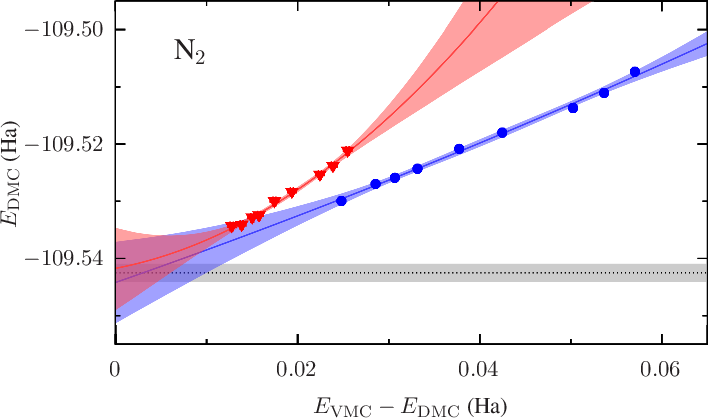} \quad
  \includegraphics[width=0.46\columnwidth]{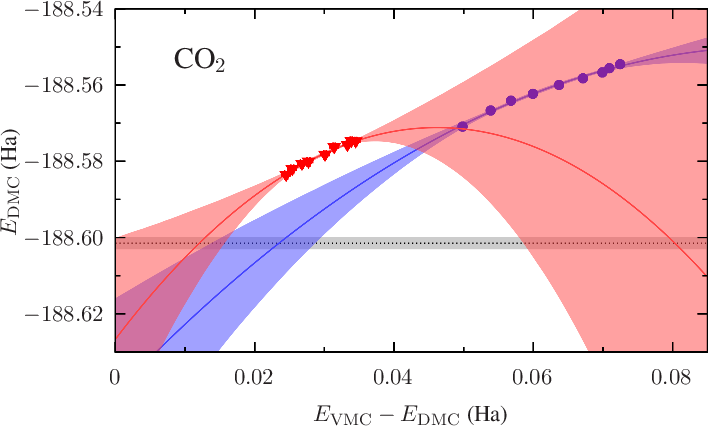}
  \caption{
    Non-backflow (blue circles) and backflow (red triangles) DMC
    energy as a function of the corresponding VMC-DMC energy
    difference for the atoms and molecules considered in this work.
    Mean values of the fits to the data are shown as lines, and the
    translucent areas around them represent 95.5\% (two-sigma)
    confidence intervals.
    Also shown in each plot is the relevant benchmark energy as a
    dotted line with a shaded area of $\pm 1$ kcal/mol around it.
  }
  \label{fig:altfit}
\end{figure}

We have applied DMC variance extrapolation to our non-backflow and
backflow data using the same fitting methodology we have used for
\textsc{xspot} and in Fig.\ \ref{fig:altfit} we show the resulting
fits.
The energy data exhibit significant curvature, justifying the need for
the quadratic fitting function, and out quasirandom fluctuation
analysis produces values of $\alpha$ of the same order of magnitude as
with \textsc{xspot}.
We find that, for the most part, DMC variance extrapolation works, but
yields much larger uncertainties than \textsc{xspot}.
This is likely the case due to the data points being less
evenly-spaced and often clustering together, resulting in several fits
with non-monotonic mean values.
It is noteworthy that the non-backflow and backflow curves do not line
up in any of the cases, which one might expect them to since the
extrapolation method is in principle agnostic to the specifics of the
wave function.

\FloatBarrier
\clearpage

\begin{figure}[!hbt]
  \centering
  \includegraphics[width=0.55\columnwidth]{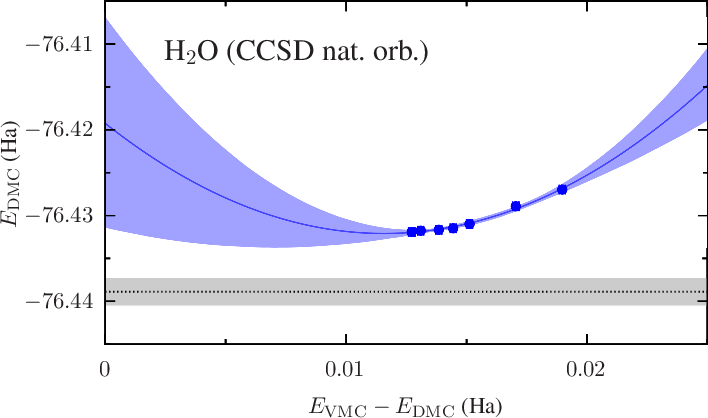}
  \caption{
    DMC energy as a function of the VMC-DMC energy
    difference for H$_2$O using CCSD natural orbitals.
    The mean value of the fit to the data is shown as a line, and the
    translucent area around it represents the 95.5\% (two-sigma)
    confidence interval.
    Also shown is the benchmark energy as a dotted line with a shaded
    area of $\pm 1$ kcal/mol around it.
  }
  \label{fig:h2o_nat_altfit}
\end{figure}

We have also applied DMC variance extrapolation to H$_2$O with natural
orbitals, see Fig.\ \ref{fig:h2o_nat_altfit}, to see if the
$w$-independent character of this approach gets around the
orbital-basis exhaustion problem encountered by \textsc{xspot}, as
mentioned above.
We find that, in contrast with \textsc{xspot} which fails to produce
an intersection as shown in Fig.\ 6 of the manuscript, DMC variance
extrapolation does yield an energy estimate, but it is affected by a
significant uncertainty and misses the benchmark by just over three
sigma, which is overall a worse-quality result than when Hartree-Fock
orbitals are used.

While DMC variance extrapolation has its merits, we believe that the
\textsc{xspot} method we present in our manuscript is a superior
option when applicable.

\FloatBarrier

\end{document}